\RequirePackage{fix-cm}
\documentclass{svjour3}                     % onecolumn (standard format)
\smartqed  % flush right qed marks, e.g. at end of proof
\usepackage{graphicx}
\usepackage{rotating}
\usepackage{longtable}
\usepackage{hyperref}
\usepackage{multirow}
\usepackage{natbib}
\usepackage{textcomp}
\usepackage{amssymb}
\usepackage{wrapfig}
\usepackage{lscape}
\usepackage{rotating}
\usepackage{epstopdf}
\usepackage{hyperref}
\usepackage{multirow}
\usepackage{natbib}
\usepackage{textcomp}
\usepackage{amssymb}
\usepackage{wrapfig}
\usepackage{lscape}
\usepackage{rotating}
\usepackage{epstopdf}

\begin{document}

\title{Software Ecosystems: A Tertiary Study and a Thematic Model}


\author{Paulo Malcher \and
        Olavo Barbosa \and 
        Davi Viana \and \newline
        Rodrigo Santos
}


\institute{Paulo Malcher \at
             % first address \\
             % Tel.: +123-45-678910\\
             % Fax: +123-45-678910\\
              \email{malcher@edu.unirio.br}           %  \\
         \and
         Olavo Barbosa \at
         \email{oalpb@cesar.school}
         \and
         Davi Viana \at
         \email{davi.viana@ufma.br}
         \and
         Rodrigo Santos \at
         \email{rps@uniriotec.br}
}

\date{Received: date / Accepted: date}

\maketitle

\begin{abstract}
A software ecosystem (SECO) is an interaction, communication, cooperation, and synergy among a set of players. Depending on the actors' type of interaction with others, each one can play a different role. These interactions provide a set of positive relationships (symbiosis) between actors who work together around a common technology platform or a service. SECO has been explored in several studies, some related to their general characteristics and others focusing on a specific topic (e.g., requirements, governance, open-source, mobile). There are many literature reviews of different natures (e.g., systematic literature reviews and systematic mapping studies). This study presents the status of the SECO field motivated by analyzing several secondary studies published over the years. To do so, we conducted a tertiary study. From an initial set of 518 studies on the subject, we selected 22 studies. We identified the theoretical foundations used by researchers and their influences and relationships with other ecosystems. We performed a thematic synthesis and identified one high-order theme, 5 themes, 10 subthemes, and 206 categories. As a result, we proposed a thematic model for SECO containing five themes, namely: social, technical, business, management, and an evaluation theme named Software Ecosystems Assessment Models (SEAM). Our main conclusion is that relationships between SECO themes should not be seen in isolation, and it must be interpreted in a holistic approach, given the number of implications to other themes mainly related to the distinction of governance and management activities in the SECO interactions. Finally, this work provides an overview of the field and points out areas for future research, such as the need of SECO community to further investigate the results from other ecosystems, mainly from the Digital Ecosystem and Digital Business Ecosystem communities.


\end{abstract}

\section{Introduction}
\label{sec:introduction}

During the last years, several concepts related to the reduction of organizational boundaries, collaboration of several agents (internal and external to the organization) to build products around a common platform have been emerged and contributed to the establishment of Software Ecosystems (SECO). These concepts instigate change in the software business  models, such as the emergence of new domains of application and the possibilities of different types of users’ experiences. \cite{SantosWerner2012} report that the pressure for innovations to be exercised at the boundaries of the organization enabled the transition from software product line to SECO. SECO was considered a new approach to software development \citep{Santos2020}. Moreover, SECO has been becoming a model more common to software development, on different agents cooperate around a shared platform \citep{axelsson2016}.

According to \cite{barbosa2013} and \cite{manikas2016revisiting}, the research about SECO started in 2003 with Messerschmitt and Szyperski’s book \citep{messerschmitt2003software}. Since that, Literature presents different definitions of SECO \citep{lungu2008towards, bosch2009, jansen2009sense}. On Scopus digital library , which indexes several other databases, more than 900 published works, quote the term “software ecosystem” since 2004.  \cite{vegendla2018systematic} affirm that SECO attracted attention in both academics and professionals. \cite{jansen2020focus} reports that the concept of SECO caused a large impact on business world and researches. Still for the author, during a short time, scientists and companies conceptualized and implemented SECO in a manner that affected significantly the society and software industry. In this regard, \cite{manikas2016revisiting} proposed an actualization of SECO definition in 2016. According to the author, SECO is as a group of agents and artifacts, internal and external to an organization or community, which exchange resources and information centered in a common technology platform. Furthermore, the author concludes in his work that the SECO field has “maturity signs”, but argues that is still quite immature, because lacks of theory, methods, or specific tolls for problems of SECO. \cite{garcia2018} affirm that is possible to study SECO by analyzing several aspects and subdomains. They argue that despite the majority studies have been focused on SECO in a general form, others focus on an aspect or specific subdomain.

The studies about SECO in literature can be classified in primary and secondary studies, being the last type focused on classifying and characterizing research works published in the field \citep{kitchenham2009}. Several secondary studies were conducted in SECO context in the last years. These studies included general reviews and different topics, for example mobile, open-source, governance, requirement, quality, and health. According to \cite{khan2019}, secondary studies are well established in software engineering (SE) and are useful for providing an overview and/or mapping of published work. Thus, it becomes important to characterize the common sense about SECO and its practice more broadly than is possible with secondary studies, involving its various types, aspects, and subdomains. Therefore, This tertiary study aims to investigate the extensive research done on SECO after almost two decades that the term has been introduced into SE, through an increasing number of primary studies and the conduct of many secondary studies. Although SECO has been the topics of several secondary studies \citep{barbosa2013, axelsson2016, garcia2018, vegendla2018systematic}, it was missing a specific tertiary study that characterized these studies.

Tertiary studies provide a evaluation of secondary studies \citep{kitchenham2010, verner2012systematic}. For \cite{khan2019}, with the increase of the secondary studies in SE, there is a need for more studies of tertiary research, because these studies play an important role to future researches and in providing evidence about the impact of secondary studies, as well as on areas in increasing on SE. In the case of this study, the main goal is to present the current condition of the researches in SECO as evidenced for secondary studies in this field. Moreover, it seeks to present a valuation regarding the characteristics, quality, and limitations of this study. 

This work contributes to provide a categorized catalog of secondary studies lead by the contexts of SECO and to present a view of research panoramas and SECO practice in software engineer. Moreover, by performing a thematic synthesis we proposed a thematic model based on five themes, as namely, social, technical, business, management and Software Ecosystems Assessment Models (SEAM). The ISO/IEC 38500 governance model inspired the identification and classification of SEAM theme. This theme aims to evaluate the current and future use of technical aspects of a SECO. Besides, it monitors conformance to policies, and performance against plans related to all themes, business, technical and social themes.

In addition, another theme called  management is the linkage between all technical, business, social, and SEAM. These results came after we performed a thematic synthesis in which we identified one high-order theme, six themes, ten subthemes, and dozens of categories in the SECO context. The rest of this article is structured as follows. The theoretical foundation is presented in Section \ref{sec:background}. Section \ref{sec:researchmethod} reports the tertiary study design, including the research method followed to conduct the study. Section \ref{sec:results} presents the results. Section \ref{sec:discussion} discusses the answers to the specific research questions. Section \ref{sec:Limitations} discusses the limitations and validity of the study. Finally, the conclusions and future work are presented in Section \ref{sec:conclusion}.

\section{Background}
\label{sec:background}

SECO studies have become an active research field and defined for the emergence of new models of software architecture, a collaboration of developers, and open business \citep{hanssendyba2012}. In a traditional approach, software development is realized in a monolithic and in closed environment. According to \cite{hanssendyba2012}, SECO approach represents a radical leap in how SE is being done. In the SECO context, development is becoming an open process in a complex distributed environment. In turn, \cite{axelsson2016} explain that SECO is based on an open environment and has a range of agents of different organizations, interacting around an open or semi-opened platform and prospering among themselves in a cooperatively and competitively manner.

The term ``ecosystems" originated from ecology, being defined as a community of living organisms and non-living components, as well the relations with each other and with the environment, interacting as a system \citep{smith2012elements}.  \cite{barbosa2013} summarize the major part of the terms and acronym used to refer SECO as Software Ecosystems or SECO, Digital (Business) Ecosystem, Mobile Learning Ecosystem or Mobile Ecosystem, Free and Open Source Software (FOSS) Ecosystem or Open Ecosystems. \cite{jansen2009business} define SECO as a set of agents working as a unit, interacting as a distributed market among software and services, including also the relations with these agents. Such relations are frequently supported for the existence of a common technology platform or for a common market, based on an exchange of information, resources, and artifacts. SECO are a set of software solutions that permit automating activities and transitions done for agents in the social ecosystem or from associated businesses and organizations, which offer these solutions \citep{bosch2009}. \cite{bosch2009} argues that enterprise development has opened their architectures in a way so that external developers collaborate, bringing up this new concept: software solutions, enterprises and developers benefiting from a common platform.

Alternatively, \cite{santos2010revisiting} present three key elements in SECO. The first element is the software: it exists as a common technology platform, central software technology, software solutions, and the platform of software or line product. The second one is transaction, taken as an understanding that embracing the model of gain, reward or also benefits non-financial. Finally, the third element refers to relationships; they are the connections among the agents from the elements, software, and transactions. Additionally, there are two essential concepts in SECO approach. First, it is a common interest in central software technology, and then it is the concept of network of organizations, that relates to symbiotic aspects, of joint evolution, commercial or technical, where each one of the agents can be benefit from participation in SECO \citep{hanssen2012, hanssendyba2012}. In this context, \cite{hanssen2012} identified three key-roles by considering the different relationships that the organization can have with a platform, namely: (i) keystone: an organization or group that leads the platform development; (ii) end-users: represent who needs the platform to realize its own business; and (iii) external developers: who uses the platform as a base to produce solutions or related services. 

Thus, it is possible to establish a view of SECO in three dimensions, initially distinguished by \cite{campbell2010} and evolved by \cite{Santos2011proposal}, divided into technical, business, and social. On the technical dimension, the goal is to open the platforms and development of external agents. Three factors guided this dimension, the development infrastructure, governance, and requirements management. On the business dimension, the goal is related to collaboration, competitiveness, the stakeholders’ expectations, and the productivity impact of the environment. Three factors guide the business dimension: environment view, innovation, and strategic planning. Finally, the social dimension focuses on organization opening as a whole, platform community, and development collaboration. Three factors guide the social dimension: utility, promotion and knowledge gain.

\section{Research Method}
\label{sec:researchmethod}

Evidence is knowledge obtained from findings derived from data analysis obtained from observational or experimental procedures that are potentially repeatable and meet the currently accepted standards of design, execution, and analysis \citep{kitchenham2002}. \cite{kitchenham2002} also state that evidence refers to knowledge obtained from findings derived from interpretation of data produced by observational or experimental procedures that are potentially repeatable and that meet accepted standards of design, execution, and analysis. Correspondingly, Evidence-Based Software Engineering (EBSE) aims to apply an evidence-based approach to SE research and practice \citep{kitchenham2007}. According to \cite{cruzes2011a}, two literature review methods has been largely implemented as a key element of EBSE: Systematic Literature Reviews (SLR) and Systematic Mapping Study (SMS). They concisely summarise the best available evidence that uses explicit and rigorous methods to identify, critically appraise, and synthesize relevant studies on a particular topic.

The tertiary study presented in this article follows the same methodology of SLR and was developed following the guidelines proposed by \cite{kitchenham2007} and \cite{petersen2015guidelines}. According to \cite{curcio2019}, these guidelines help researchers design, conduct and evaluate empirical research. Although they have different methodology proposals, the guidelines are similar. In addition, they are guidelines widely accepted in the EBSE community and used in highly cited tertiary studies in the literature, such as \citep{kitchenham2010, cruzes2011a, curcio2019, khan2019}, which also inspired this study. 

According to \cite{kitchenham2007}, the following phases are proposed: planning, conducting, and reporting the results. The planning phase has the objective of defining the justification, scope, restrictions, and all the research strategy to be used in the review. The conduction phase includes the following activities: searching for articles, selecting them, articles quality evaluation, and then extracting data, analyzing data, and synthesing data to answer the research questions. Finally, the results report phase consists of the summarization of results, based on the analysis and synthesis of the data conducted after the extraction phase. Similar to secondary studies, the procedures for conducting tertiary studies are defined in a review protocol. A review protocol specifies the methods that will be used to perform a specific systematic review and reduces the possibility of bias by the researcher \citep{kitchenham2004}. The steps of each phase of the review are detailed in the following subsections.

\subsection{Objective and Research Questions}

The study’s objetive is to analyze the characteristics, quality, and coverage and evaluate the potential impact of the results of secondary studies conducted in the SECO research field on the progress of research and practice in SE. To do so, we intended to answer the research questions defined in Table \ref{tab:researchquestions}.

\begin{table}[h]
    \caption{Research questions.}
    \label{tab:researchquestions}
    \centering
    \begin{tabular}{|l|l|}
    \hline
\textbf{\textbf{ID}} & \textbf{\textbf{Research questions}}\\
\hline
RQ1 & What are the general and methodological characteristics of the secondary studies on SECO? \\ \hline
RQ1.1 &	What are the demographics data of secondary studies on SECO? \\ \hline
RQ1.2 &	What are the types (SMS or SLR) of secondary studies on SECO? \\ \hline
RQ1.3 &	What are the publication venues of secondary studies on SECO? \\ \hline
RQ1.4 &	How is the quality of the secondary studies on SECO?\\ \hline
RQ1.5 &	Which research topics are being addressed by secondary studies on SECO? \\ \hline
RQ1.6 &	What are the limitations and/or threats to the validity of secondary studies on SECO? \\ \hline
RQ2	& What are the characteristics of the theoretical backgrounds and the findings of secondary studies on SECO? \\ \hline
RQ2.1 &	Which theoretical foundations are adopted by secondary studies on SECO? \\ \hline
RQ2.2 &	Which themes have been addressed by secondary studies on SECO? \\ \hline
RQ2.3 &	What are the gaps of research topics pointed by secondary studies on SECO? \\ \hline
    \end{tabular}
\end{table}

%\begin{tabular}{p{6.4cm}p{1.3cm}}

%%\begin{table}[h!]
%%caption{Research questions.}
%%\label{tab:researchquestions}
%%\begin{tabular}{|l|l|}
%%\hline
%%\multicolumn{1}{|c|}{\textbf{ID}} & \multicolumn{1}{c|}{\textbf{Research questions}} \\ \hline
%%RQ1 & What are the general and methodological characteristics of the secondary studies on SECO? \\ \hline
%%RQ1.1 &	What are the demographics data of secondary studies on SECO? \\ \hline
%%RQ1.2 &	What are the types (SMS or SLR) of secondary studies on SECO? \\ \hline
%%RQ1.3 &	What are the publication venues of secondary studies on SECO? \\ \hline
%%RQ1.4 &	How is the quality of the secondary studies on SECO?\\ \hline
%%RQ1.5 &	Which research topics are being addressed by secondary studies on SECO? \\ \hline
%%RQ1.6 &	What are the limitations and/or threats to the validity of secondary studies on SECO? \\ \hline
%%RQ2	& What are the characteristics of the theoretical backgrounds and the findings of secondary studies on SECO? \\ \hline
%%RQ2.1 &	Which theoretical foundations are adopted by secondary studies on SECO? \\ \hline
%%RQ2.2 &	Which themes have been addressed by secondary studies on SECO? \\ \hline
%%RQ2.3 &	What are the gaps of research topics pointed by secondary studies on SECO? \\ \hline
%%\end{tabular}
%%\end{table}

\subsection{Research Strategy}

The search strategy used involved automatic searches in digital libraries. The selected search sources were ACM Digital Library\footnote{https://dl.acm.org/}, Elsevier ScienceDirect\footnote{https://www.sciencedirect.com/}, Engineering Village\footnote{https://www.engineeringvillage.com/}, IEEEXplore Digital Library\footnote{https://ieeexplore.ieee.org/Xplore/}, Scopus\footnote{https://www.scopus.com/}, SpringerLink\footnote{https://link.springer.com/}, and Web of Science\footnote{https://webofknowledge.com/}. All the sources were used in several studies, such as Kitchenham et al. (2010), Cruzes and Dybå (2011), Khan et al. (2019), and Curcio et al. (2019). The method used to perform searches in digital libraries followed the strategy of initially identifying keywords and synonyms from terms present in research questions. Therefore, search for terms used in relevant articles that have conducted secondary studies in SECO and generate a Search String using boolean operators such as OR and AND to incorporate the terms. Finally, perform searches on selected search sources (application of search string). The search string has been changed slightly according to the format required by each search source. Table \ref{tab:searchstring} provides the search string.

\begin{table}[h]
\centering
\caption{The search string.}
\label{tab:searchstring}
\begin{tabular}{|l|}
\hline
\multicolumn{1}{|c|}{\textbf{The search string}} \\ 
\hline
(``software ecosystems" OR ``software ecosystem") AND (``review of studies" OR ``structured review" OR \\ ``systematic review" OR ``literature review" OR ``literature analysis" OR ``systematic literature review" OR \\``systematic map" OR ``systematic mapping" OR ``mapping study" OR ``longitudinal study" OR ``scoping study")\\ \hline
\end{tabular}
\end{table}

\subsection{Study Selection}

All returned studies were inserted into the Mendeley tool\footnote{https://www.mendeley.com}, and then all duplicates were removed in order to optimize the selection process. According to \cite{kitchenham2007}, for the selection of the studies returned, inclusion criteria (IC) and exclusion criteria (EC) must be defined and applied (Table \ref{tab:inclusioncriteria} and Table \ref{tab:exclusioncriteria}). They are used to exclude studies that are not relevant to answer the research questions. The study selection involved four researchers with experience in conducting and publishing secondary studies (three of them with more than 10 years of experience). Two researchers conducted the selection phase of studies by reading the title, abstract, and keywords of all studies and applying the inclusion and exclusion criteria. Studies were included if they covered all IC and excluded if they were within at least one of the EC. In case of disagreement in the application of the IC/EC, another researcher decided the conflicts. We could not obtain some studies and we had to request them directly from the authors.

\begin{table}[h]
\centering
\caption{Inclusion criteria.}
\label{tab:inclusioncriteria}
\begin{tabular}{|l|l|}
\hline
\multicolumn{1}{|c|}{\textbf{ID}} & \multicolumn{1}{c|}{\textbf{Criteria}} \\ \hline
IC1 & \begin{tabular}[c]{@{}l@{}}The study that described secondary studies strategies such as SLR Longitudinal Studies or SMS \\ (or Scoping Studies) on the SECO research topic.\end{tabular}\\ \hline
IC2	& The study is written in English.\\ \hline
IC3	& The study is peer-reviewed.\\ \hline
\end{tabular}
\end{table}

\begin{table}[h!]
\centering
\caption{Exclusion criteria.}
\label{tab:exclusioncriteria}
\begin{tabular}{|l|l|}
\hline
\multicolumn{1}{|c|}{\textbf{ID}} & \multicolumn{1}{c|}{\textbf{Criteria}} \\ \hline
EC1	& The study does not satisfy the inclusion criteria.\\ \hline
EC2 & The same study is found in more than one search engine. In this case, just one study was considered.\\ \hline
EC3 & Duplicate studies reporting similar results. In this case, only the most complete study was considered.\\ \hline
EC4	& Inaccessible studies by free download or that were not sent after our e-mail request.\\ \hline
EC5 & The study is an MSc or PhD thesis. \\ \hline
\end{tabular}
\end{table}

\subsection{Quality Assessment}

The quality assessment of studies included is essential in secondary and tertiary studies \citep{kitchenham2015evidence, zhou2015quality}. The quality assessment determines to what extent the results of an empirical study are valid and bias-free. The reliability of results and conclusions drawn from systematic reviews and some types of systematic mapping studies is associated with their quality \citep{kitchenham2015evidence}. It is important to evaluate the secondary studies in order to analyze how these studies are designed despite the fact that quality assessment has already been performed in primary studies. 

In this tertiary study, an adaptation of the quality criteria used by \cite{kitchenham2015evidence} was used. These criteria were defined by the Centre for Reviews and Disseminations (CRD) of York University and became known as the Database of Abstracts of Reviews of Effects (DARE). The criteria used are shown in Table \ref{tab:qualityassessmentcriteria}.

\begin{table}[h!]
\centering
\caption{Quality assessment criteria. Adapted from  \cite{kitchenham2015evidence}.}
\label{tab:qualityassessmentcriteria}
\begin{tabular}{|l|l|}
\hline
\multicolumn{1}{|c|}{\textbf{ID}} & \multicolumn{1}{c|}{\textbf{Criteria}} \\ \hline
QAC1 & Did the reviewers report the bias associated to the process? \\ \hline
QAC2 & Did the reviewers report the internal validity criteria associated to the process? \\ \hline
QAC3 & Did the reviewers report the external validity associated to the process? \\ \hline
QAC4 & Did the reviewers report the methods used for data extraction and synthesis? \\ \hline
QAC5 & Are the review’s inclusion and exclusion criteria described? \\ \hline
QAC6 & Did the reviewers assess the quality/validity of the included studies? \\ \hline
QAC7 & Was the search adequate? \\ \hline
QAC8 &	Are sufficient details about the individual included studies presented? \\ \hline
QAC9 & Were the included studies synthesized? \\ \hline
\end{tabular}
\end{table}

\begin{figure}[h]
	\centerline{\includegraphics[scale=1.0]{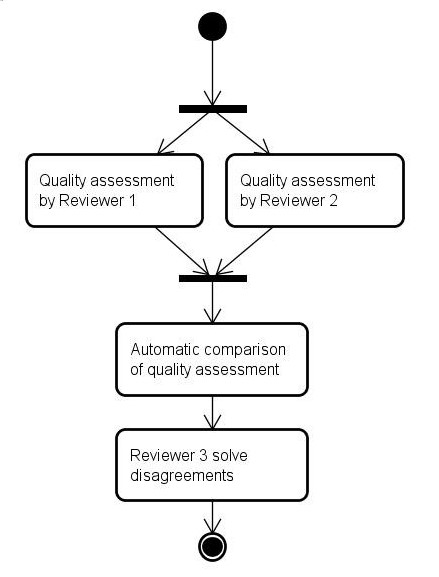}}
	\caption{Quality assessment process. 
	\label{fig:qualityassessmentprocess}}
\end{figure}

We defined three levels of scoring for each criteria: Yes = 1.0, Partially = 0.5 and No, or unknown = 0. Table \ref{tab:qualityassessmentitems} lists the criteria and scoring guidelines used for evaluation. Two researchers initially conducted the evaluation process independently and the results were compared in order to resolve any points of disagreement. In case of disagreement, a third researcher (with more experience) resolved the issue. The process is illustrated in Figure \ref{fig:qualityassessmentprocess}.

\begin{table}[h]
\centering
\caption{Quality assessment items. Adapted from \cite{kitchenham2015evidence}.}
\label{tab:qualityassessmentitems}
\begin{tabular}{|l|l|l|l|}
\hline
\multicolumn{1}{|c|}{\textbf{Criteria}} & \multicolumn{1}{c|}{\textbf{Yes}} &
\multicolumn{1}{|c|}{\textbf{Partially}} & \multicolumn{1}{c|}{\textbf{No}} \\ \hline
QAC1 & \begin{tabular}[c]{@{}l@{}}There are preventative steps taken to\\ minimize bias and errors in\\ the study selection process \end{tabular} & The criteria are implicit &	\begin{tabular}[c]{@{}l@{}} Criteria are not defined and\\ cannot be readily inferred \end{tabular}\\ \hline
QAC2 & Authors report explicitly & The criteria are implicit & \begin{tabular}[c]{@{}l@{}} Criteria are not defined and\\ cannot be readily inferred \end{tabular}\\ \hline
QAC3 & Authors report explicitly & The criteria are implicit & \begin{tabular}[c]{@{}l@{}} Criteria are not defined and\\ cannot be readily inferred \end{tabular}\\ \hline   
QAC4 & \begin{tabular}[c]{@{}l@{}} There are preventative steps taken to\\ minimize bias and errors in the data\\ extraction process and synthesis \end{tabular}  & The criteria are implicit &	\begin{tabular}[c]{@{}l@{}} The inclusion criteria are not\\ defined and cannot be readily\\ inferred \end{tabular}\\ \hline
QAC5 & \begin{tabular}[c]{@{}l@{}} The inclusion criteria are explicitly\\ defined in the study \end{tabular} & \begin{tabular}[c]{@{}l@{}}  The inclusion criteria are\\ implicit \end{tabular} & \begin{tabular}[c]{@{}l@{}}  The inclusion criteria are not\\ defined and cannot be readily\\ inferred \end{tabular} \\ \hline
QAC6 & \begin{tabular}[c]{@{}l@{}} The authors have explicitly defined\\ quality criteria and extracted them from\\ each primary study, and there are\\ preventative steps to minimize bias and\\ errors in the quality assessment process \end{tabular} & \begin{tabular}[c]{@{}l@{}} The research question\\ involves quality issues that\\ are addressed by the study \end{tabular} & \begin{tabular}[c]{@{}l@{}} No explicit quality assessment\\ of individual primary studies\\ has been attempted, or authors\\ have defined quality criteria\\ but not used the criteria \end{tabular} \\ \hline
QAC7 & \begin{tabular}[c]{@{}l@{}}The authors have either searched 4 or\\ more digital libraries and included\\ additional search strategies or identified\\ and referenced all journals addressing\\ the topic of interest \end{tabular} & \begin{tabular}[c]{@{}l@{}} The authors have searched\\ 3 or 4 digital libraries with\\ no extra search strategies,\\ or searched a defined but\\ restricted set of\\ journals and conference\\ proceedings \end{tabular} & \begin{tabular}[c]{@{}l@{}} The authors have searched up\\ to 2 digital libraries or an\\ extremely restricted set of\\ journals \end{tabular} \\ \hline
QAC8 & \begin{tabular}[c]{@{}l@{}}Information is presented about each\\ primary study \end{tabular} & \begin{tabular}[c]{@{}l@{}} Only summary information\\ is presented about studies \end{tabular} & \begin{tabular}[c]{@{}l@{}} The results of the individual\\ studies are not specified \end{tabular} \\ \hline
QAC9 & \begin{tabular}[c]{@{}l@{}}The authors have explicitly stated the\\ synthesis method and provided a\\ reference to that method \end{tabular} & \begin{tabular}[c]{@{}l@{}} The authors have stated\\ their synthesis method \end{tabular} & \begin{tabular}[c]{@{}l@{}} The authors do not describe\\ their synthesis method \end{tabular} \\ \hline
\end{tabular}
\end{table}

\subsection{Data extraction and synthesis}

The data extraction step in a tertiary study comprises the design of forms for accurately recording the information obtained in secondary studies \citep{kitchenham2007}. The data extraction process of this study was carried out from the complete reading of the included studies and the use of a spreadsheet to store the data. The objective of the data synthesis is to analyze and summarize the evidence of the included studies to answer the research questions of this study. To do so, we used descriptive analysis \citep{cruzes2011b}. The process of reading and extracting data followed the following strategy: Two researchers extracted data from the studies by conducting numerous meetings to analyze data extractions elucidate doubts, agreements, and disagreements.

After screening the studies and quality assessment, one of the researchers extracted data correlated to RQ1. Thus, tables, graphs, diagrams were built in order to summarize the evidence from the secondary studies. In other words, for all these studies included, the following data were extracted: title, authors, country/location, digital library, year of publication, type of secondary study, type of publication, publication venue, number of primary studies included, quality indicators and guidelines used, research topics, limitations, theoretical foundations, themes, and gaps. To answer RQ2 the research synthesis approach was performed, which will be detailed in Subsection \ref{researchsynthesisapproach}.

\subsubsection{Research synthesis approach}
\label{researchsynthesisapproach}

After the first researcher was done, the second one started analyzing secondary studies starting from the oldest to newest one. By reading from the oldest to the newest one, it was easier to identify the cross-references between the secondary studies mainly related to the theoretical foundations used by the articles. Thus, it was possible to identify all the knowledge used from secondary studies, despite the methodological content and the findings. In this context, \cite{cruzes2011b} state that a research synthesis can be considered a collective term used to summarize, integrate, combine, and compare the findings of primary studies. We used this approach by setting the secondary studies as the objects of analysis. Thus, for the synthesis of these secondary studies, thematic analysis was used as a method for identifying, analyzing and reporting patterns (or themes) from data to help us in drawing conclusions. It was used as a method of both interpretative and aggregative method for the main findings of the studies mainly regarding answering the RQ2.

Regarding thematic analysis, \cite{cruzes2011b} propose five steps for thematic synthesis. The first consists of extracting data from the primary studies. The second involves coding data by identifying and coding interesting concepts, categories, findings, and results. The third translates codes into themes, sub-themes, and higher-order themes. The fourth step creates a model of higher-order themes by exploring relationships between themes and create a model of higher-order themes. Finally, the last one assesses the trustworthiness of the synthesis and interpretations leading up to the thematic synthesis. Once the unit of analysis of this review is the set of secondary studies, the process of coding was not performed because regardless of secondary study’s authors following this thematic analysis or not, each secondary study already done some kind of synthesis, being either aggregative type (where the thematic analysis fits) or interpretive. Thus, the coding in the context of secondary studies are much closer to the concepts and results found where we derive the categories.

\begin{figure}[h]
	\centerline{\includegraphics[scale=0.6]{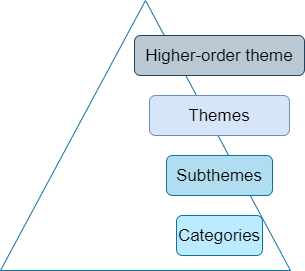}}
	\caption{Hierarchy of elements of the thematic synthesis. Adapted from the works of Connelly and Peltzer [40], Cruzes and Dybå [39].
	\label{fig:HierarchyOfElements}}
\end{figure}

For developing themes in the literature, we used an adaption of the works of the \cite{cruzes2011b} and \cite{connelly2016}. Unlike Cruzes and Dybå, Connelly and Peltzer explicitly recognize the use of categories of the process pyramid elements of developing themes. In this regard, \cite{connelly2016} state an interesting advice regarding themes creation. They state that when developing a theme, each word in the theme should be carefully chosen. The meaning of each word should be clear and appropriate for the purpose of the study. For instance, the word social could be problematic because there is no way of knowing what the term stands alone as a meaning. The same could happen with other terms or keywords such as business, architectural, platform, or technical. Thus, the elements considered in this study are for example higher-order themes can be composed of themes; themes from subthemes; subthemes from categories, and categories are identified from the analysis of qualitative data. Figure \ref{fig:HierarchyOfElements} illustrates the hierarchy of elements used in the thematic analysis.

\begin{figure}[h]
	\centerline{\includegraphics[scale=0.33]{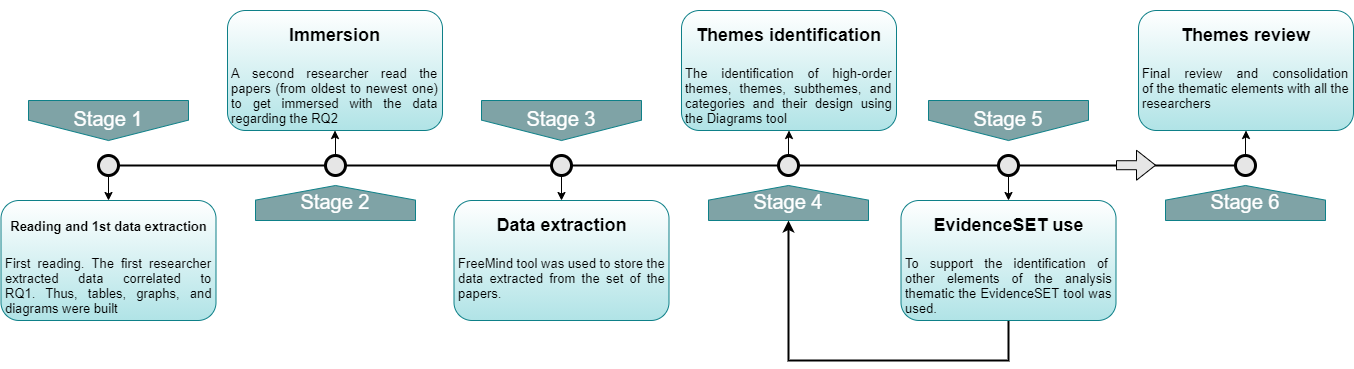}}
	\caption{Process of data extraction and themes identification and review.
	\label{fig:ProcessOfExtracting}}
\end{figure}

To support the identification of the elements of this hierarchy, we used tools such as EvidneceSET\footnote{https://evidenceset.com.br/}, Diagrams\footnote{https://app.diagrams.net/} and FreeMind \footnote{https://freemind.sourceforge.net/}. According to \cite{barbosa2017}, EvidenceSET is a web-based tool to support the identification of research themes from a set of primary or secondary studies. EvidenceSET provides a set of views based on the number of frequencies that words appear in a set of studies. On this subject, \cite{kitchenham2015evidence} state that thematic analysis is more interpretative than content analysis. \cite{kitchenham2015evidence} argues that if the thematic analysis stops identifying first-order codes, its findings remain closely related to the original data. Nevertheless, the process used by this study went beyond this approach. We also used part of the processes defined by \cite{guest2012} that state that thematic analyses may include: comparing code frequencies, identifying code co-occurrence, and graphically displaying relationships between codes within the data set. With the reference of aggregative methods, quantitative content analysis is a kind of aggregative method that also involves the process of counting the number of times that some specific words are mentioned in text \citep{kitchenham2015evidence}. Thus, we used EvidenceSET because its views provide a way to display relationships between concepts within the dataset, word frequencies, and word co-occurrence. Diagrams are a tool to design several kinds of visualizations from the business to the technical perspectives. Finally, FreeMind is an open-source tool to create mental maps. We used it to store the collected data regarding the study connections, namely: similarities concerning SECO definitions; the use of the concept of SECO 3D; research agenda; research topics; and research designs. Nonetheless, before using any tool and reducing the bias of being suggested by them, we followed the manual identification of the elements of the hierarchy of Figure \ref{fig:HierarchyOfElements}. Figure \ref{fig:ProcessOfExtracting} illustrates the process of the data extraction used in this tertiary study.

\section{Results}
\label{sec:results}

This section describes the secondary studies selection, which followed the procedures defined in the review protocol and detailed in the previous section. 518 studies were returned, and 22 secondary studies were selected. The selection process is presented in Figure \ref{fig:SelectionProcess}. The complete list of included secondary studies can be seen in Appendix \ref{SecStudies}. The secondary studies are referred to as S1-S22.

\begin{figure}[h]
	\centerline{\includegraphics[scale=0.65]{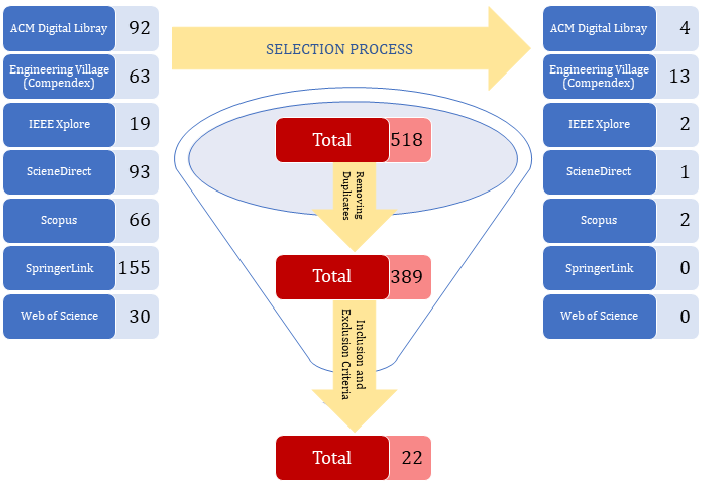}}
	\caption{Secondary studies selection process.
	\label{fig:SelectionProcess}}
\end{figure}

The following subsections present the answers to the research questions, defined through data extraction and synthesis. In addition, the results of the quality evaluation performed in each included secondary study were taken into consideration.

\subsection{RQ1: What are the general and methodological characteristics of the secondary studies on SECO?}

This RQ identified general characteristics of the secondary studies included in our research. One of them is the presentation of demography data that allowed us to visualize tendencies and gaps for the research field. Additionally, we used some kind of classification to identify studies according to their goal and type of research questions. Thus, after analyzing the secondary studies, we identified some topics covered by them, as namely: general, measures, OSS, quality, open innovation, mobile, health, requirements engineering, governance, partnership models, and software architecture. It is important to highlight that all of them are related to SECO. For instance, health is not covered alone; it is addressed in a SECO context. Therefore, we classified a study as Governance topic when governance is the main concern of the research question of a given study. The same can be applied to all identified topics. In special, the general topic means that the research question used by the authors is   an open question, that is, it is a wider scope of study, regardless whether it is an SMS or SLR. Regarding the other information, we identified the principal’s publications venues, years of publications, number of primary studies included country of origin of the main author, and the most frequent synonyms used to designate SECO. Table \ref{tab:GeneralCharacteristics} presents general characteristics of studies. Some of these characteristics will be better detailed in other research questions in this work.

\begin{table}[h!]
\centering
\caption{General characteristics of secondary studies}
\label{tab:GeneralCharacteristics}
\begin{tabular}{|l|l|l|l|l|l|l|}
\hline
\multicolumn{1}{|c|}{\textbf{ID}} & \multicolumn{1}{c|}{\textbf{Year}} & \multicolumn{1}{c|}{\textbf{Venue*}} & \multicolumn{1}{c|}{\textbf{Type**}} & \multicolumn{1}{c|}{\textbf{Number of primary studies included}} & \multicolumn{1}{c|}{\textbf{Country}} & \multicolumn{1}{c|}{\textbf{Topic}} \\ \hline
S1 & 2013 & B & SMS & 44 & Brazil & General \\ \hline
S2 & 2013 & J &	SLR & 90 & Denmark & General  \\ \hline
S3 & 2014 & C &	SMS & 34 & Sweden & measures  \\ \hline
S4 & 2014 &	C &	SLR & 17 & Spain & OSS and Quality  \\ \hline
S5 & 2015 &	C &	SMS & 260 &	Sweden & Open innovation  \\ \hline
S6 & 2015 &	C &	SMS & 28 & Brazil &	Mobile  \\ \hline
S7 & 2015 &	C &	SLR & 38 & Finland & SECO health  \\ \hline
S8 & 2016 &	J &	SLR & 231 &	Denmark & General  \\ \hline
S9 & 2016 &	J &	SMS & 6 & Sweden & Quality  \\ \hline
S10 & 2017 & C & SLR & 23 & Brazil & SECO health  \\ \hline
S11 & 2017 & J & SMS & 82 &	Spain & OSS  \\ \hline
S12 & 2018 & J & SMS & 44 &	Norway & RE  \\ \hline
S13 & 2018 & C & SMS & 13 &	Spain & General  \\ \hline
S14 & 2018 & C & SMS & 12 &	Brazil & SECO health  \\ \hline
S15	& 2018 & C & SLR & 89 &	Brazil & \begin{tabular}[c]{@{}l@{}} Governance and\\ SECO health \end{tabular}  \\ \hline
S16	& 2019 & C & SMS & 23 &	Brazil & Partnership models  \\ \hline
S17	& 2019 & W & SMS & 63 &	Brazil & Mobile  \\ \hline
S18	& 2019 & W & SLR & 15 &	Brazil & Mobile  \\ \hline
S19	& 2019 & J & SLR & 37 &	Spain & General  \\ \hline
S20	& 2019 & J & SMS & 23 &	Spain & General  \\ \hline
S21	& 2019 & C & SMS & 44 &	Brazil & \begin{tabular}[c]{@{}l@{}}  Software\\ Architecture \end{tabular}  \\ \hline
S22	& 2019 & C & SMS & 49 &	Brazil & Open innovation  \\ \hline
\end{tabular}
\raggedright{\emph{*Publication Type: B: Book Chapter / C: Conference / J: Journal / W: Workshop}}\\
\raggedright{\emph{**Secondary Study Type: SLR: Systematic Literature Review / SMS: Systematic Mapping Study}}
\end{table}

Table \ref{tab:AcronymsAndSynonyms} presents the principal searched terms in secondary studies included related to SECO topics and themes (acronyms and synonyms). These terms were gathered from research strings of each study and it shows the way that researchers have been collected primarily in SECO context.

\begin{table}[h!]
\centering
\caption{Acronyms and synonyms for SECO}
\label{tab:AcronymsAndSynonyms}
\begin{tabular}{|l|l|}
\hline
\multicolumn{1}{|c|}{\textbf{Acronyms and synonyms for SECO}} & \multicolumn{1}{c|}{\textbf{Study ID}}  \\ \hline
Software ecosystem*	& \begin{tabular}[c]{@{}l@{}} [S1] [S2] [S3] [S5] [S6] [S7] [S8] [S9] [S10] [S12] \\ {[S13] [S14] [S15] [S16] [S17] [S18] [S21]} \end{tabular} \\ \hline
SECO &	[S13] [S14] [S17] [S18] [S19] \\ \hline
Open Source Ecosystem &	[S4] [S5] [S10] [S11] \\ \hline
Software supply network* & [S1] [S3] [S11] \\ \hline
Software supply industry &	[S1] [S11] \\ \hline
Software platform & [S16] [S17] \\ \hline
OSS Ecosystem & [S4] [S11] \\ \hline
FOSS Ecosystem & [S4] [S11] \\ \hline
FLOSS Ecosystem & [S4] [S11] \\ \hline
Free Software Ecosystem & [S4] [S11] \\ \hline
Libre Software Ecosystem & [S4] [S11] \\ \hline
Mobile software ecosystem &	[S17] [S18] \\ \hline
MSECO & [S17] [S18] \\ \hline
Mobile ecosystem* & [S3] \\ \hline
Software vendor* & [S1] \\ \hline
Digital ecosystem* & [S3] \\ \hline
Embedded ecosystem & [S5] \\ \hline
Product development ecosystem & [S5] \\ \hline
Software product-line/product line/product line ecosystem & [S5] \\ \hline
Software third party ecosystem & [S5] \\ \hline
Technological ecosystem* & [S13] \\ \hline
Information ecosystem* & [S13] \\ \hline
ERP ecosystem* & [S13] \\ \hline
Open ecosystem* & [S13] \\ \hline
Learning ecosystem* & [S13] \\ \hline
Software digital ecosystem* & [S14] \\ \hline
\end{tabular}
\raggedright{\emph{The use of ``*" in the end of some words means that we used plural variations of the terms. For instance, Software ecosystem* means both “Software ecosystem” and “Software ecosystems”.}}\\
\end{table}

\subsubsection{RQ1.1: What are the demographics data of secondary studies on SECO?}

\begin{figure}[h]
	\centerline{\includegraphics[scale=0.6]{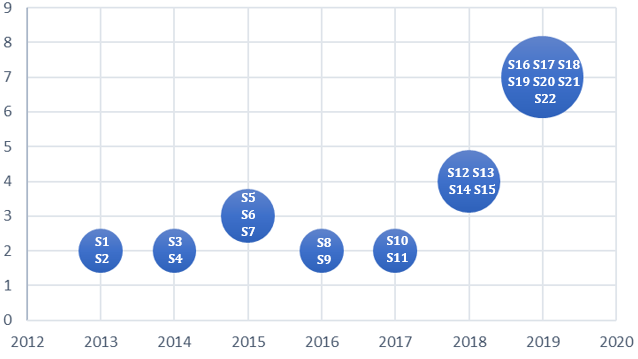}}
	\caption{Distribution of secondary studies by year.
	\label{fig:DistributionByYear}}
\end{figure}

RQ1.1 identified demographic data of secondary studies. We notice the consistency in publications of secondary studies between the years of 2013 and 2017 (2 or 3 per year). The first secondary study was published in 2012 and, in 2019, 7 secondary studies related to a topic of research in SECO, as can be observed in Figure \ref{fig:DistributionByYear}. In 2018, there was an increase on the quantity of studies (4 studies) and, in 2019, 7 studies was published. This increase in publications corroborates with \cite{khan2019}, which claim that there is an increase of number of secondary studies on SE in general. RQ1.1 too was related to countries identified in the studies. 

\begin{figure}[h]
	\centerline{\includegraphics[scale=0.09]{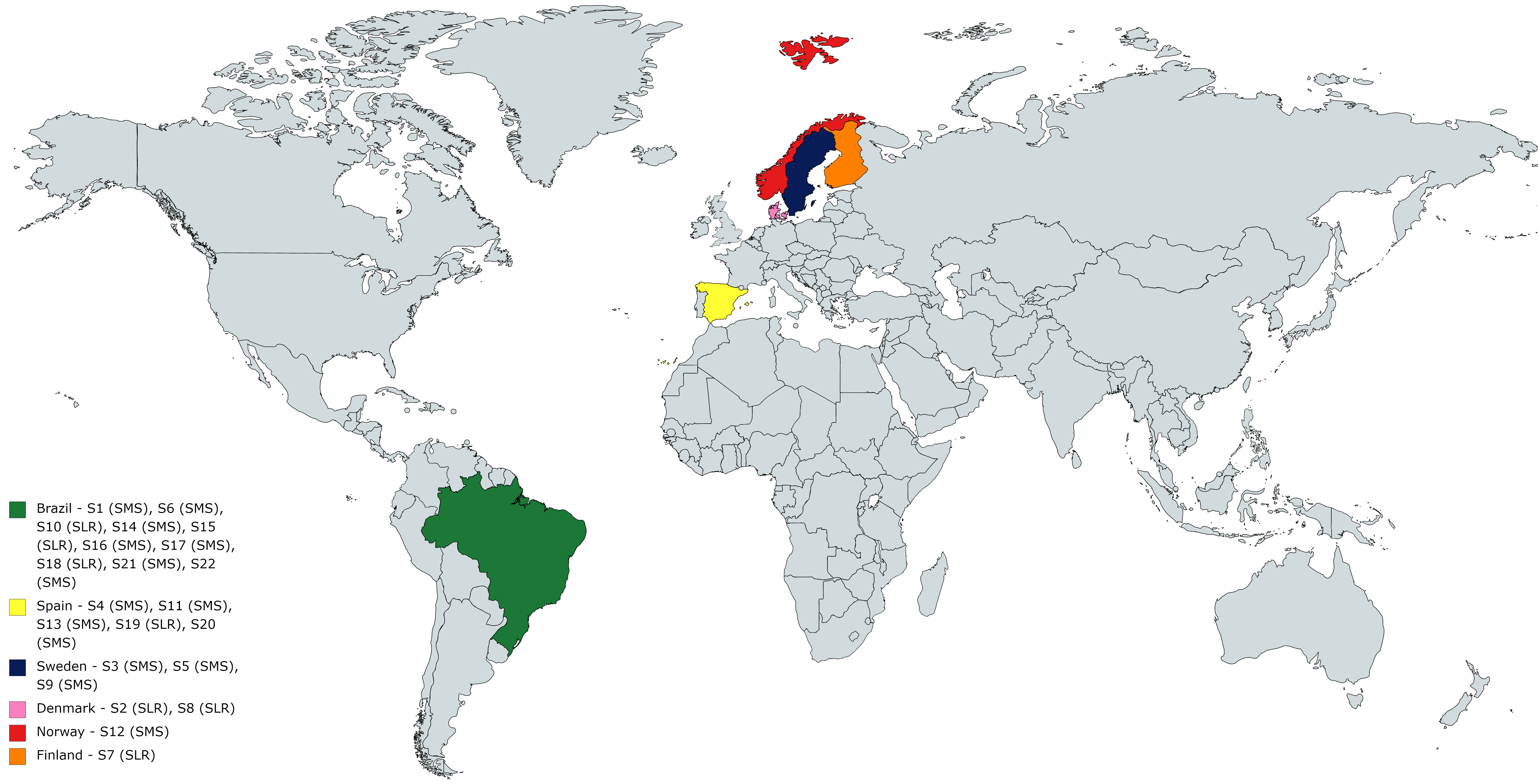}}
	\caption{Distribution of secondary studies by country.
	\label{fig:DistributionByCountry.png}}
\end{figure}

The results (Figure \ref{fig:DistributionByCountry.png}) points out a significant number of secondary studies from Brazil (10) and Spain (5). It is important highlight that we considered the nationality of the first author to define the country of each study. Four studies have presented information about the countries of the studies. S5 identifies United States of America as the country with more publications about this field. For S11, Europe is the continent more dominant, followed by North America. S17 presents a collaboration network of researchers of mobile software ecosystem (MSECO) and researchers from Brazil and Finland have centrality. S20 intended to analyze the European context and highlighted Italy, Germany, Austria and Greece as the countries with more publications. However, we observed that the huge part of secondary studies included (18) did not mention the countries of primary studies and presents the nationality of studies can allow viewing collaboration network and the community know the countries that have more published in the field.

\subsubsection{RQ1.2: What are the types (SMS or SLR) of secondary studies on SECO?}

\begin{figure}[h]
	\centerline{\includegraphics[scale=0.5]{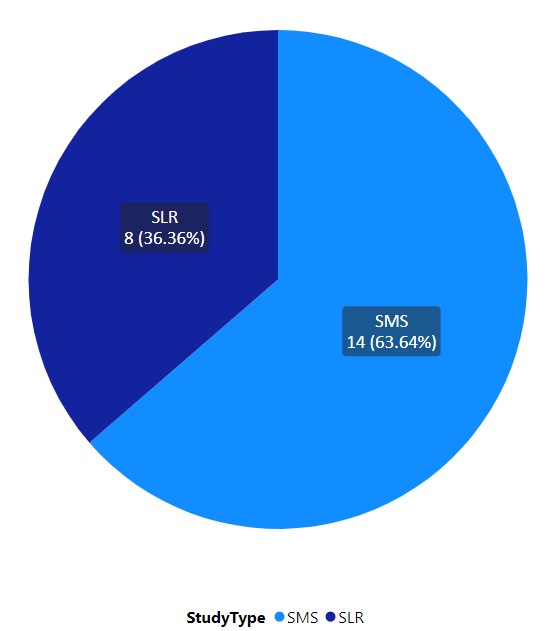}}
	\caption{Secondary studies selection process
	\label{fig:DistributionOfSecondaryStudiesType}}
\end{figure}

We observed that the number of SMS (14) exceeds the number of SLR (8) (Figure \ref{fig:DistributionOfSecondaryStudiesType}). It may be justified due to the emergence of SECO field over years. We analyzed and categorized according to the guidelines for both SLR by \cite{kitchenham2007} or SMS by \cite{petersen2008systematic}.

\subsubsection{RQ1.3: What are the publication venues of secondary studies on SECO?}

We identified the main publication venues (conferences, books, workshops and journals) of secondary studies in SECO. We noted that conferences such as the International Conference on Software Business (ICSOB) and the International Conference on Enterprise Information Systems (ICEIS) concentrated efforts of researchers of this area with four and three publications, respectively. In addition, The Journal of Systems and Software prominently featured as the journal with four publications. It is worth mentioning that events directly related to SECO have emerged in the last decades, such as International Workshop on Software Ecosystems (IWSECO) and Workshop on Software Engineering for Systems-of-Systems and Workshop on Distributed Software Development, Software Ecosystems and Systems-of-Systems (SESoS/WDES). Among 22 secondary studies included, 6 came from journals, 13 from conferences, 2 were published in workshops, and 1 is a book chapter. Table \ref{tab:PublicationVenues} presents the complete list of publications venues and Figure 8 presents the number and percentage of publication type.

\begin{table}[h]
    \caption{Publication venues for the included secondary studies.}
    \label{tab:PublicationVenues}
    \centering
    \begin{tabular}{|l|l|l|}
    \hline
\textbf{Venue} & \textbf{Public. Type} & \textbf{Study ID}\\
\hline
%\begin{tabular}[c]{\centering@{}l@{}} Publication\\ Type \end{tabular}
Int. Conference on Software Business (ICSOB) & C & [S3] [S5] [S7] [S21] \\ \hline
The Journal of Systems and Software (JSS) &	J &	[S2] [S8] [S9] \\ \hline
Int. Conference on Enterprise Information Systems (ICEIS) &	C &	[S15] [S16] \\ \hline
\begin{tabular}[c]{@{}l@{}} Software Ecosystems - Analyzing and\\ Managing Business Networks in the Software \end{tabular} &	B &	[S1] \\ \hline
Int. Conference on Software Technologies (ICSOFT) &	C &	[S4] \\ \hline
Int. Computers, Software \& Applications Conference (COMPSAC) & C & [S6] \\ \hline
Brazilian Symposium on Software Engineering (SBES) & C & [S10] \\ \hline
Information and Software Technology (IST) & J &	[S11] \\ \hline
Journal of Information Technology Research (JITR) &	J &	[S12] \\ \hline
\begin{tabular}[c]{@{}l@{}} Int. Conference on Technological Ecosystems for\\ Enhancing Multiculturality (TEEM) \end{tabular} & C & [S13] \\ \hline
\begin{tabular}[c]{@{}l@{}} Euromicro Conference on Software Engineering and\\ Advanced Applications (SEAA) \end{tabular} & C & [S16] \\ \hline
\begin{tabular}[c]{@{}l@{}} Int. Workshop on Software Engineering for Systems-of-Systems and\\ Workshop on Distributed Software Development, Software Ecosystems and\\ Systems-of-Systems (SESoS/WDES) \end{tabular} & W & [S17] \\ \hline
\begin{tabular}[c]{@{}l@{}} Int. Workshop on Cooperative and Human Aspects of Software\\ Engineering (CHASE) \end{tabular} & W & [S18] \\ \hline
Sensors & J & [S19] \\ \hline
Journal of Medical Systems (JMS) & J & [S20] \\ \hline
\begin{tabular}[c]{@{}l@{}} Int. Conference on Product-Focused Software\\ Process Improvement (ICPFSPI) \end{tabular} &	C & [S22] \\ \hline
\end{tabular}
\raggedright{\emph{*Publication Type: B: Book Chapter / C: Conference / J: Journal / W: Workshop.}}\\
\end{table}

\begin{figure}[h]
	\centerline{\includegraphics[scale=0.5]{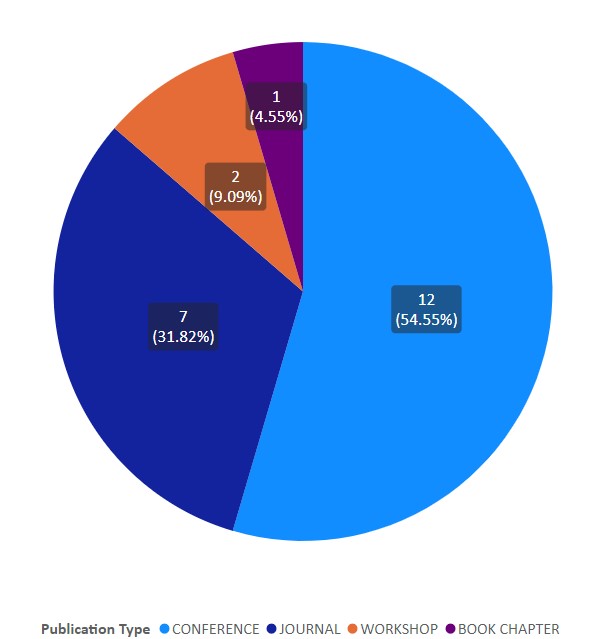}}
	\caption{Number and percentage of publication type.
	\label{fig:NumberAndPercentageOfPublicationType}}
\end{figure}

\subsubsection{RQ1.4: How is the quality of the secondary studies on SECO?}

\begin{figure}[h]
	\centerline{\includegraphics[scale=0.6]{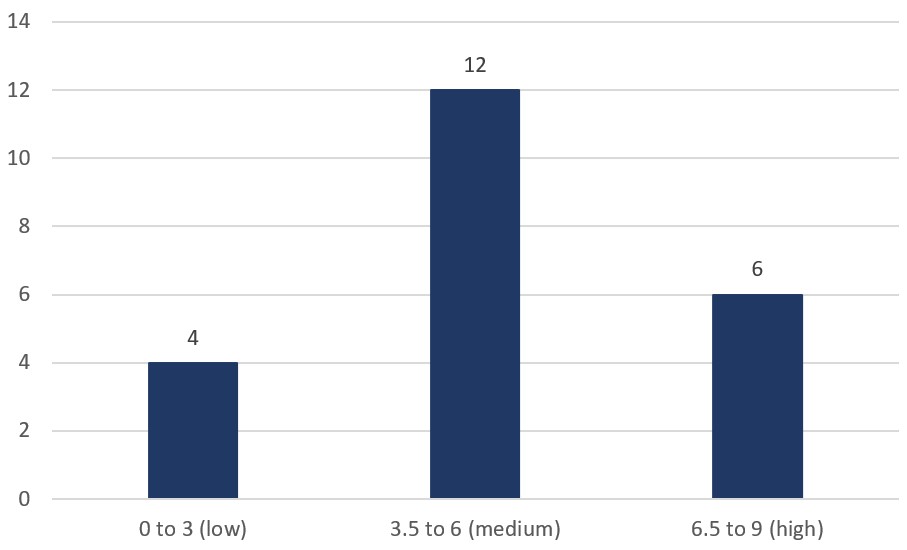}}
	\caption{Quality ranges of secondary studies.
	\label{fig:QualityRanges}}
\end{figure}

Related to the quality criteria, it was possible to notice that the SLR studies were better elaborated compared to SMS. Moreover, the quality of secondary studies published in journals was higher than the quality of secondary studies published in conferences. The frequency of score 1 in each quality criteria to secondary studies published in journals is higher than secondary studies published in conferences, as in the relation between SLR and SLM. The relation of average scores between both characteristics can be viewed in Figure \ref{fig:QualityRanges}. Only six secondary studies presented quality criteria and evaluated the quality of primary studies. Among these six, three were studies based in SLR (37,5\% out of total of SLR) and four were SMS (28,6\% out of total of SMS).  All studies were appraised by a quality assessment done by two researchers and validated for a third researcher that solved possible inconsistencies. Based on the quality criteria presented above, we reached the final score of each publication.

The total score of quality that we assigned to each secondary study varies from 0 to 3 (low), 3.5 to 6 (average) and 6.5 to 9 (high). The distribution of quality scores is showed in Figure \ref{fig:QualityRanges}. It indicates how a quality varies among the secondary studies included. The results of the analyses shows that in general the secondary studies published have average quality (the average quality score is 5.5 in a total of 9.0). If we analyze individual results, it is possible to visualize the number of studies with low quality is  low. 4 out of 22 studies had score equal or under to 3. The most of studies were concentrated in classifications of average or high quality with 12 studies in range from 3.5 to 6.0, and six had score equal or over to 6.5. No study was scored with maximum score (9 points). \cite{petersen2008systematic} explains that this is expected in SMS, but can bring problems related to accuracy and relevance of primary studies in LSR. Beyond that, we noticed that half of studies used objective criteria to evaluate the quality of studies did not present any guideline to define from these (Table \ref{tab:QualityIndicators}).

\begin{figure}[h]
	\centerline{\includegraphics[scale=0.4]{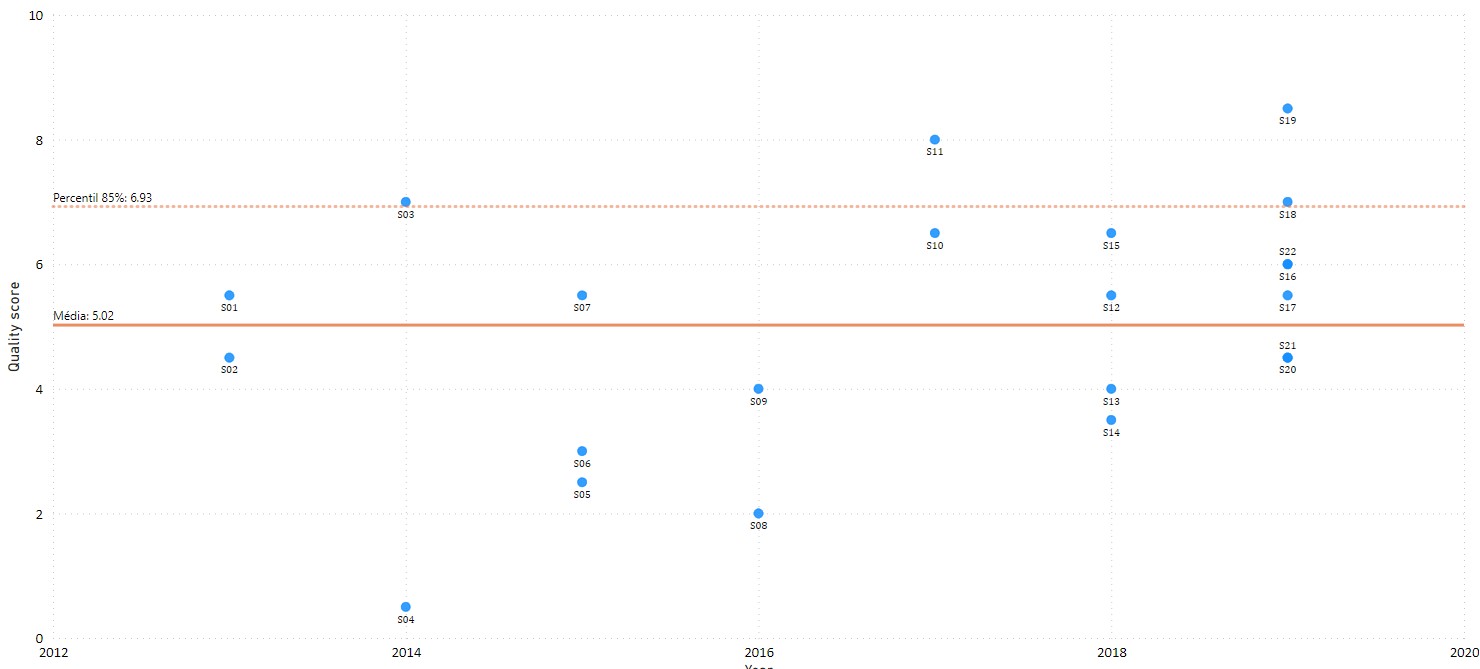}}
	\caption{Distribution of quality scores by year.  
	\label{fig:QualityScoresByYear}}
\end{figure}

Figure \ref{fig:QualityScoresByYear} shows a quality variation from secondary studies published from 2013 to 2019. It can be observed through the graphic of dispersal that the most quality scores are concentrated on the upper right quadrant. The vertical blue line represents the average of the quality analysis. The gray line, in turn, represents the 85\% percentile of the quality analysis values. This means that approximately 85\% of the studies have a score above the average and, in particular, two of them have a so high score, the studies S11 and S19. There are more secondary studies with average or high-quality score in the last years, as shown by the trend line of the graphic. Table \ref{tab:QualityIndicators} shows a quality score from secondary studies included. Figure \ref{fig:NumberAndPercentageOfPublicationType} shows the number of publications, countries and quality scores over years. The increase of publications in the last two years can be observed. Besides, although we have calculated the quality average for each year, it does not mean that the indication of quality has improved or decreased over the years due to the small number of studies per year. However, there is a constancy regarding the number of countries involved in the publications.

\begin{figure}[h]
	\centerline{\includegraphics[scale=0.45]{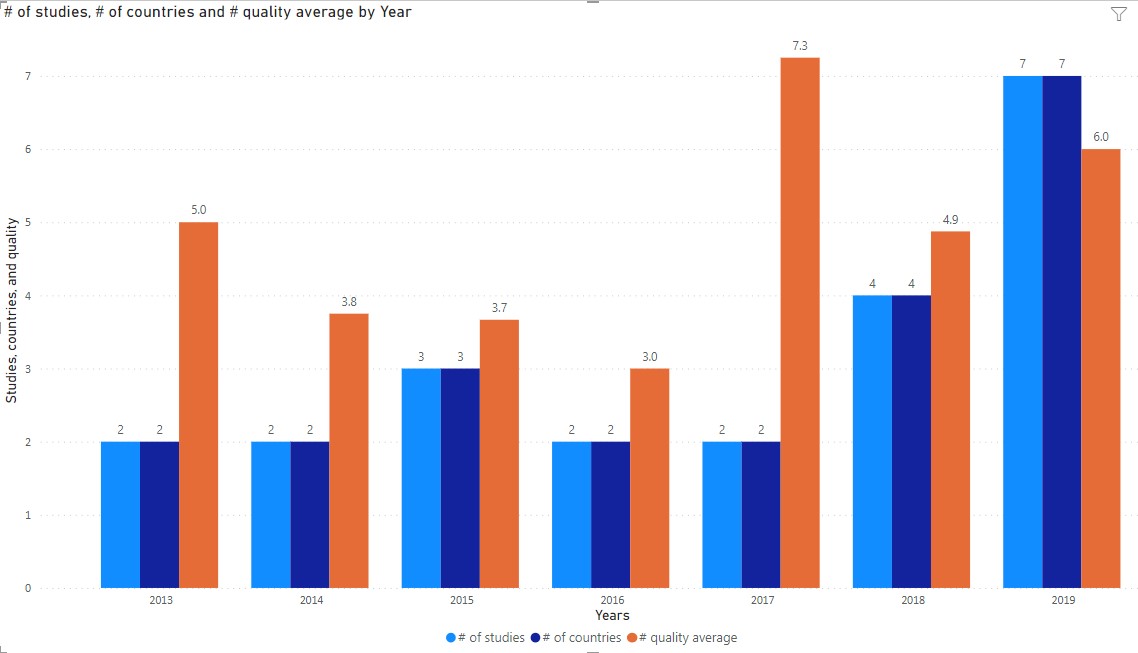}}
	\caption{Number of publications, countries and quality score over years.
	\label{fig:NumberAndPercentagePublicationType}}
\end{figure}

\begin{table}[h]
\caption{Quality indicators of secondary studies}
\label{tab:QualityIndicators}
\centering
\begin{tabular}{|l|l|l|l|l|l|}
\hline
\textbf{ID} & \textbf{Type} & \textbf{Quality Assessment?} & \textbf{Follow Guidelines?} & \textbf{Guidelines} & \textbf{Quality Score}\\ \hline
%\begin{tabular}[c]{\centering@{}l@{}} Publication\\ Type \end{tabular}
S1 & SMS & No &	Not applicable & Not applicable	& 5.5 \\ \hline
S2 & SLR & No &	Not applicable & Not applicable	& 4.5 \\ \hline
S3 & SMS & No &	Not applicable & Not applicable	& 7 \\ \hline
S4 & SLR & No &	Not applicable & Not applicable	& 0.5 \\ \hline
S5 & SMS & No &	Not applicable & Not applicable	& 2.5 \\ \hline
S6 & SMS & No &	Not applicable & Not applicable	& 3 \\ \hline
S7 & SLR & No &	Not applicable & Not applicable	& 5.5 \\ \hline
S8 & SLR & No &	Not applicable & Not applicable	& 2 \\ \hline
S9 & SMS & No &	Not applicable & Not applicable & 4 \\ \hline
S10 & SLR & Yes & No & Not applicable &	6.5 \\ \hline
S11 & SMS & No & Not applicable & Not applicable & 8 \\ \hline
S12 & SMS &	Yes & Yes &	\begin{tabular}[c]{@{}l@{}} (Nguyen-Duc et al., 2015) and\\
Dybå and Dingsøyr’s checklist \end{tabular} &	5.5 \\ \hline
S13 & SMS & Yes & No & Not applicable &	4 \\ \hline
S14	& SMS &	No & Not applicable & Not applicable & 3.5 \\ \hline
S15	& SLR &	No & Not applicable & Not applicable & 6.5 \\ \hline
S16	& SMS &	No & Not applicable & Not applicable & 6 \\ \hline
S17	& SMS &	No & Not applicable & Not applicable & 5.5 \\ \hline
S18	& SLR &	Yes & No &	Not applicable & 7 \\ \hline
S19	& SLR &	Yes & Yes &	\begin{tabular}[c]{@{}l@{}} Kitchenham and\\ Charters (2007) \end{tabular} & 8.5 \\ \hline
S20	& SMS &	Yes & No & Not applicable & 4.5 \\ \hline
S21	& SMS &	No & Not applicable & Not applicable & 4.5 \\ \hline
S22	& SMS &	No & Not applicable & Not applicable & 6 \\ \hline
\end{tabular}
\end{table}

\subsubsection{RQ1.5: Which research topics are being addressed by secondary studies on SECO?}

Related to researched topics, we noticed that works, which analyze SECO generally and other that focus on specific topics. All research topics defined are directly related to SECO and were defined based primarily on the objectives and research questions of the included secondary studies. For example, the research topic health was assigned to a study when its primary objective was related to investigating the health of SECO. Therefore, a study was classified with the research topic governance when governance, in the context of SECO, is the central concern of the research question of a given study. The same can be applied to all the other identified research topics. In particular, the general research topic means that the research question used by the authors is an open question, i.e., it is a broader scope of study and sought to investigate SECO generally. We also observed that some studies focused on more than one research topic (S4 and S15). 

Another highlight was the fact that six of secondary studies were realized about SECO in a general context: S1 (2013), S2 (2013), S8 (2016), S13 (2018), S19 (2019), and S20 (2019). The quantity of publications of studies can demonstrate emergent aspect of the field, the update of concepts and, finally, its expansion to other domains. Besides, 16 articles are focused in 18 specifics research  topics in context of SECO. These articles analyze a specific type of SECO mobile [S6, S17, S18] or open source software (OSS) [S4, S11]. Alternatively, a specific topic related to SECO, as health [S7, S10, S14, and S15], open innovation [S5, S22], quality [S4, S9], governance [S15], requirements engineering [S12], partnership models [S16], software architecture [S21] and measures [S3]. 

\begin{figure}[h]
	\centerline{\includegraphics[scale=0.27]{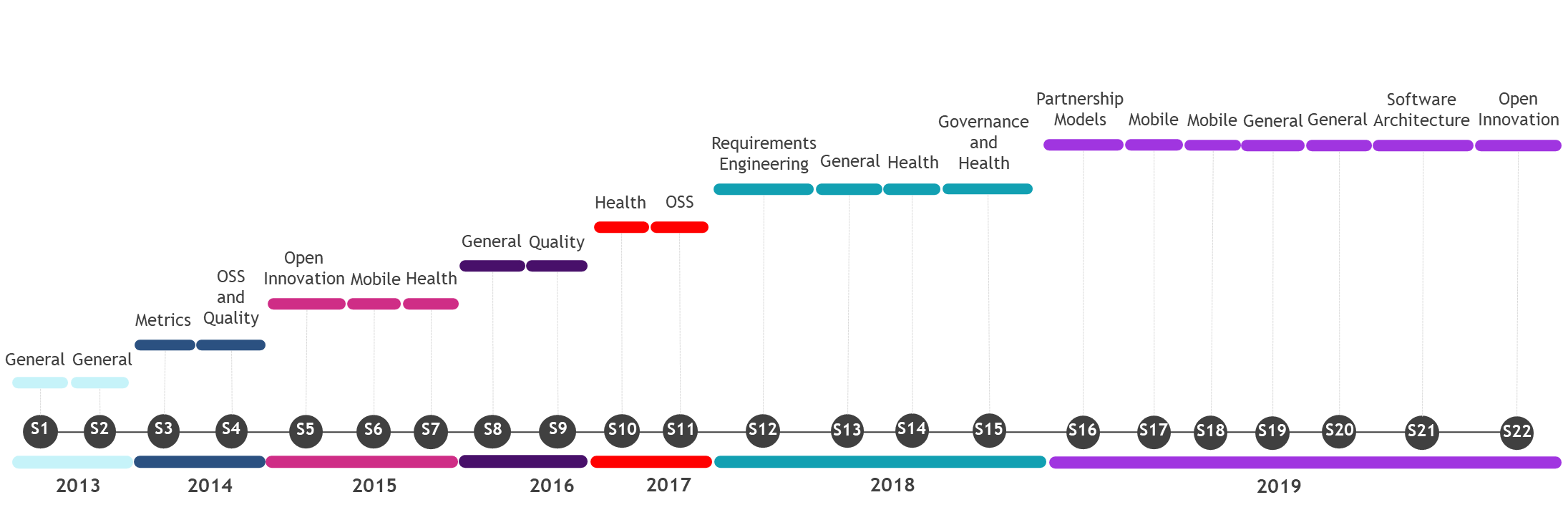}}
	\caption{Distribution of secondary studies by topic of research over the years.
	\label{fig:TopicOfResearchOverTheYears}}
\end{figure}

The research topics related to SECO can be visualized through the years in Figure \ref{fig:TopicOfResearchOverTheYears}. In this way, it is possible to notice which topics were study objects during the period of 2013 to 2019. It is important to highlight that any study had been found in 2020, even investigations had been occurred in May of 2020.

\subsubsection{RQ1.6: What are the limitations and/or threats to validity of secondary studies on SECO?}

\begin{figure}[h]
	\centerline{\includegraphics[scale=0.6]{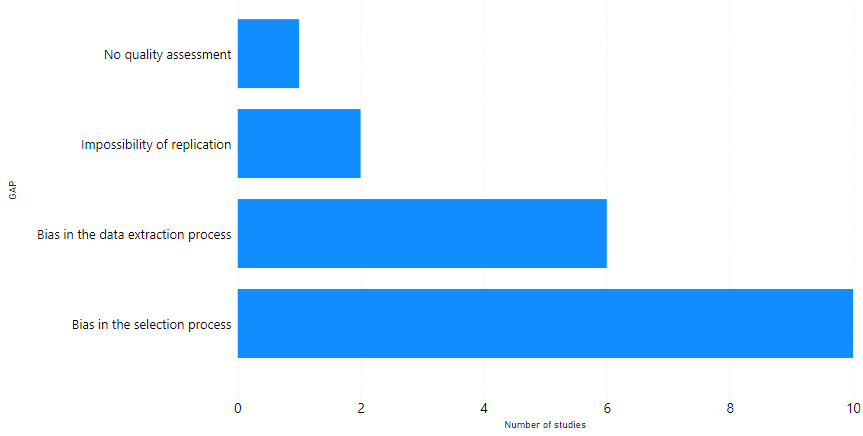}}
	\caption{Limitations and/or threats of validity of secondary studies.
	\label{fig:LimitationsThreats}}
\end{figure}

Figure \ref{fig:LimitationsThreats} presents the limitations informed on secondary studies related to the validity of its process of construction and conclusion. It is possible to understand that are inherent limitations to any secondary study, but only 11 studies quoted some limitations. The others (11) do not do reference to limitation or threat to validity. The limitation most informed is related to skew in selection process, cited in ten studies. Regarding that, it is related that relevant studies may not have been recovered in selection [S1, S4, S7, S16, S21, S22]. Besides, the definition of researches questions [S11, S19], research string [S7, S10, S19, S21] and researches source [S16, S19] also were considerate as related limitations to selection process of studies. Related to extraction tendency process, six studies highlighted as limitations, the necessity to subject analyzes on extracted data [S1, S10, S15, S19, S21]. Two studies quoted the impossible replication as a limitation [S11, S29] and only one presented as limitation the lack of quality valuation on primary studies [S2].
 
\subsection{RQ2: What are the characteristics of the theoretical backgrounds and the findings of secondary studies on SECO?}

Mapping the theoretical foundations used by researchers in the research topic, allowed us to understand important aspects in defining and orchestrating an SECO. Furthermore, the identified themes, sub-themes and categories may allow the community to visualize more broadly how SECO has been explored. Finally, gaps identified in the studies are presented and can serve as motivation for future research. 

\subsubsection{RQ2.1: Which are theoretical foundations adopted by secondary studies on SECO?}

We observed some similarities regarding the use of theoretical foundations that directed the field of SECO through the years through the analysis of secondary studies. Several concepts were introduced in the field of SECO by using established ideas, concepts, and themes, in both natural and business ecosystems. Based on the analysis of the secondary studies, the theoretical foundations of SECO come from its business, digital and biological counterparts (Table \ref{tab:TheoreticalBasis}).  They are organized as follows:  Natural ecosystems (ECO), and Business ecosystems (BECO), and Digital ecosystems (DECO), and Digital Business ecosystems (DBECO).

Business aspects, for instance, highlight the interdependence of all actors in the business environment, who “coevolve their capabilities and roles” \citep{iansiti2004keystone}. In this context, we observed some similarities, regardless of the research topic used in the secondary studies. These relationships are regarding the use of theoretical foundations that engaged the SECO field through the years. We acknowledged that although the term of SECO is already well-established research field, it has been used as a synonym for a similar term, such as ``technological ecosystems", as observed in S12, S13, S19, and S20. For instance, S20 states that collaborative partnerships of different stakeholders work around different technological platforms and these collaborative environments have evolved into what is known as technological ecosystems.

An important study about SECO is the seminal book by Messerschmitt and Szyperski \citep{messerschmitt2003software} that has been used by S1, S2, S5, S8, and S11. As identified in these studies, the concept of SECO had its first appearance in the book by Messerschmitt and Szyperski, being the birth of the research field. In relation to business, S1, S6, S7, and S11 have employed the Moore’s work \citep{moore1993predators} as theoretical basis of their findings. S1 (the first literature review in the SECO field) mapped the SECO research has been mainly inspired by studies from both business and natural ecosystems. S6 followed this approach in another SMS, that linked SECO to others kinds of ecosystems such as natural or biological ecosystems (ECO), business ecosystems (BECO), and digital ecosystems (DECO). 

S7, for instance, argue that their view of SECO follows the original work by Moore and they see SECO as an economic community.  The authors of S11 also follow S7 viewpoint when using Moore’s definition of a business ecosystem by being an economic community supported by a foundation of interacting organizations and individuals. Regarding the works of \cite{iansiti2004, iansiti2004keystone}, for instance, S1, S2, S4, S8, S9, S13, S15 and S23 have identified the studies of these authors as theoretical foundations of SECO. In S8, for instance, the identification of what SECO literature is concerned is studied. This SLR recognized health as a concept of an ecosystem, and states that this one is also a derivation from SECO from the works of Iansiti and Levien. 

Later, health topic, quality assurance or health characteristics of SECO operated as the basis of other secondary studies such as S7, S9, S10, S14, and S15. Specifically, in S15, health is also connected to the concept of governance. S15 authors updated their previous secondary study by conducting a new SLR of 89 primary studies to characterize how health is characterized in the SECO literature, having found eight definitions of the concept derived from SECO. As stated by the S15, health is the base theoretical approach to use some measure to assess the health of SECO. Regarding the establishment of the ecosystem field as a new research field, we highlight (Table \ref{tab:TheoreticalBasis}) the works of \cite{jansen2009sense}; \cite{bosch2009}; \cite{bosch2010}; \cite{campbell2010}; \cite{dhungana2010}; and \cite{manikas2013software} and all the others. All of those primary studies and the ones derived from them have been used as the conceptual basis for SECO. 

Another evidence of theoretical foundations is the use of the BECO concept of keystone. It is one of the most used terms to define an actor in the context of SECO. It comes from business ecosystems literature \citep{iansiti2004, iansiti2004keystone}. Keystone is explicitly used by S1, S2, S4, S6, S7, S9, S12 and S22. According to S1, a keystone actor is a hub responsible for creating and sharing value along ecosystem. S2 mapped some synonymous used to describe keystone role such as orchestrator, hub, and platform owner, and they define it as a company, department of a company, actor or set of actors, community or independent entity that is responsible for the well-functioning of the SECO. S4 affirms that a keystone is part of a list of measures that can be used to measure the maintenance capacity, a community characteristic of an open source assessment model. In this scenario, the keystone (or outdegree of keystone actors) does not figure out as a company, but someone who has many developers and he/she works with and plays a large role in the software ecosystem. S7 specifies keystone as an organization that provides a technological platform; sets standards and practices; and increases the value through increasing the number of contributors. S7 emphases at health of SECO and recognizes that keystones actors play an important role at moderating the factors that affects the health of ecosystem. S9 argues that keystone being as part of the set of players of the system of quality assurance in a software ecosystem.  S12 derives the concept specified by \cite{hanssendyba2012}, which states that a keystone is an organization that leads the development. \cite{hanssendyba2012}, in turn, used exactly the same concept brought by Iansiti and Levien. Finally, S22 extends the concept defined by S2 where the keystone actor is the organization that governs the evolution of the ecosystem by defining rules of access to the platform and orchestrating the creation of new solutions.

\begin{table}[h]
\caption{Theoretical basis of the reviews}
\label{tab:TheoreticalBasis}
\centering
\begin{tabular}{|l|l|l|}
\hline
\textbf{Type of ecosystem} & \textbf{\begin{tabular}[c]{@{}l@{}} Theoretical comes from the\\ following primary studies \end{tabular}} & \textbf{Cited by} \\ \hline
%\begin{tabular}[c]{\centering@{}l@{}} Publication\\ Type \end{tabular}
ECO & Tansley (1935) & [S11] [S19] \\ \hline 
BECO & Moore (1993) & [S1] [S6] [S7] [S11] \\ \hline 
BECO & Iansiti and Levien (2004a and 2004b) & [S1] [S2] [S4] [S8] [S9] [S13] [S15] [S23] \\ \hline 
BECO & Hartigh et al (2006) and Hartigh et al (2013) & [S2] [S4] [S7] [S8] [S10] [S11] [S15] \\ \hline 
DECO & Alves et al. (2009) and Stefanuto et al. (2011) & [S1] [S2] \\ \hline 
DECO & 	Stanley, J. and Briscoe, G.  (2010) & [S1] [S11] \\ \hline 
DECO & 	Briscoe and Wilde (2006) & [S1] [S6] \\ \hline 
DECO & 	Nachira et al. (2007a) & [S11] \\ \hline 
DECO & 	Briscoe (2009) & [S11] \\ \hline 
DECO & 	Briscoe and Wilde (2009) & [S1] \\ \hline 
DECO & 	Briscoe (2010) & [S2] [S8] \\ \hline 
DECO & 	Kajan et al. (2011) & [S1] [S8] [S9] \\ \hline 
DECO & 	Ostadzadeh et al. (2015) & [S13] \\ \hline 
DECO & 	Qureshi, B. (2014) & [S19] \\ \hline 
DECO & 	Iyawa et al. (2019) & [S19] \\ \hline 
DECO & 	\begin{tabular}[c]{@{}l@{}} Others primary studies published in Management\\ of Emergent Digital Ecosystems (MEDES) \end{tabular} & \begin{tabular}[c]{@{}l@{}} [S1] [S2] [S3] [S6]\\ {[S7] [S8] [S10] [S13] [S17] [S19]} \end{tabular}  \\ \hline   
SECO & Messerschmitt and Szyperski (2003) &	[S1] [S2] [S5] [S8] [S11] \\ \hline
SECO & Yu (2007) & \begin{tabular}[c]{@{}l@{}} [S1] [S2] [S4] [S6] [S8]\\ {[S11] [S12] [S15] [S17] [S18]} \end{tabular} \\ \hline 
SECO & Bosch (2009)	& [S1] [S2] [S6] [S8] [S11] [S13] [S18] \\ \hline 
SECO & Lungu (2009) & [S1] [S2] [S4] [S11] [S13] \\ \hline 
SECO & Jansen et al. (2009a) & [S1] [S2] [S6] [S11] [S12] [S15] \\ \hline 
SECO & Jansen et al. (2009b) & \begin{tabular}[c]{@{}l@{}} [S1] [S2] [S4] [S6] [S7] [S9] [S10]\\ {[S11] [S12] [S13] [S15] [S18] [S19] [S22]}  \end{tabular} \\ \hline 
SECO & Bosch and Sijtsema (2010) &	[S1] [S2] [S9] [S17] [S18] \\ \hline SECO & Dhungana et al. (2010) &	[S1] [S2] [S6] [S8] [S11] [S15] [S18] \\ \hline 
SECO & Hanssen and Dybå (2012) & [S6] [S11] [S12] [S13] [S15] \\ \hline 
SECO & Jansen and Cusumano (2012) & [S4] [S6] [S7] [S11] [S12] [S15] \\ \hline 
SECO & Christensen et al. (2014) & [S7] [S8] [S11] [S12] [S19] \\ \hline SECO & Campbell and Ahmed (2010)* &	[S1] [S2] [S9] [S12] [S17] [S18] \\ \hline 
SECO & Santos and Werner (2011a)* & [S11] \\ \hline 
SECO & Santos and Werner (2011b)* & [S2] [S7] [S15] \\ \hline 
SECO & Barbosa et al. (2013)* & [S8] [S9] [S11] [S12] [S13] [S15] \\ \hline 
SECO & Santos and Werner (2012)* & [S7] [S8] \\ \hline  
SECO & Manikas and Hansen (2013)* & [S4] [S7] [S8] [S9] [S11] [S15] [S22] \\ \hline 
DBECO &	Stanley and Briscoe (2010) & [S1] [S11]\\ \hline 
DBECO &	Nachira et al. (2007b) & [S11]\\ \hline 
\end{tabular}
\raggedright{\emph{*SECO from SECO three dimesion perspectives}}\\
\end{table}

\begin{figure}[ht]
	\centerline{\includegraphics[scale=0.2]{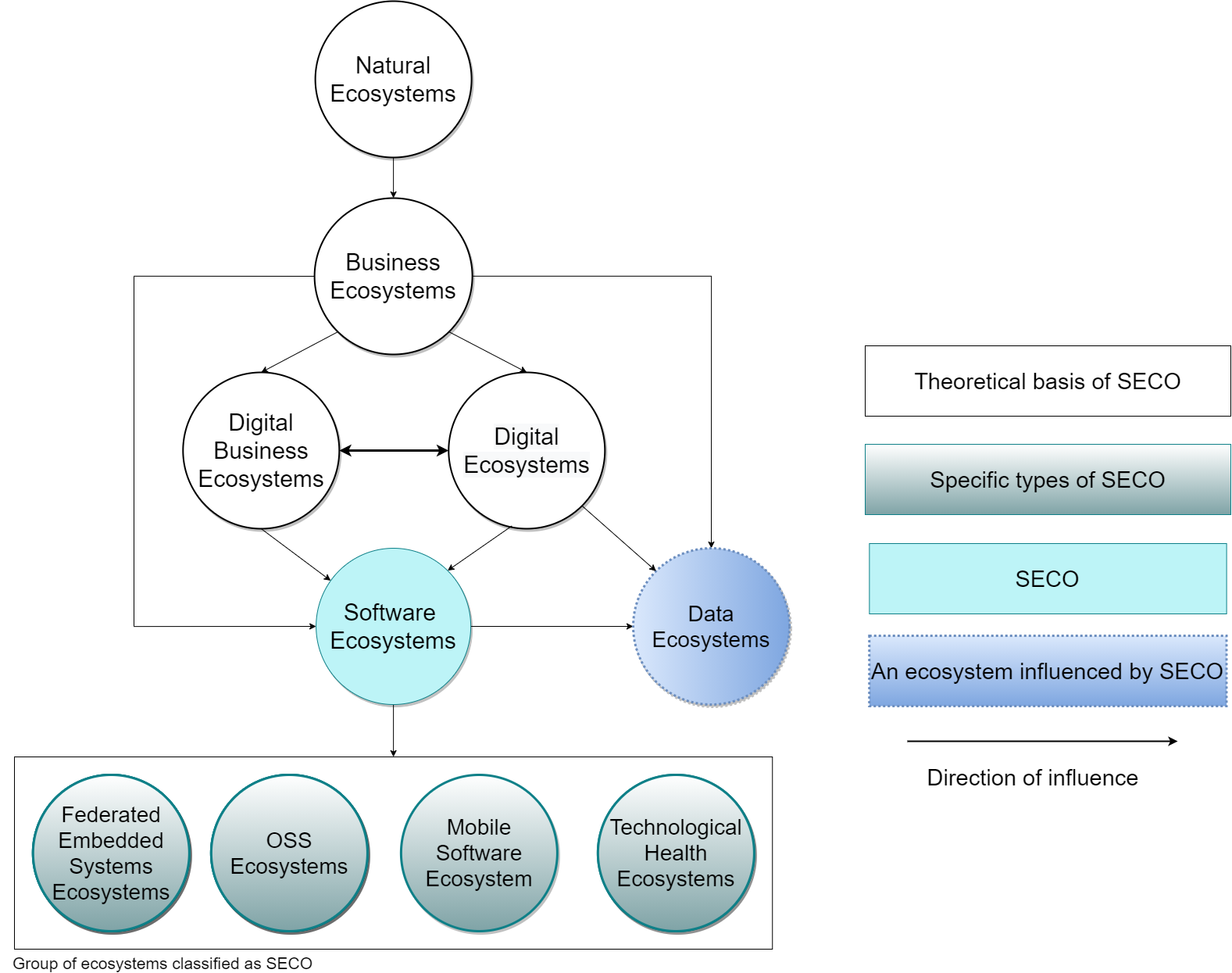}}
	\caption{Theoretical foundations of SECO and its influences.
	\label{fig:TypesOfSECO}}
\end{figure}

To illustrate the SECO in the three-dimensional perspective and its relationships, we schemed a visualization for the theoretical foundations identified by the analysis of the studies (Figure \ref{fig:TypesOfSECO}). It denotes that SECO researchers use analogies with other types of ecosystems for organizing, classifying and evaluating SECO.  The same occurs with other kinds of ecosystem as identified by \cite{nachira2007network} and \cite{Santos2011treating}. These studies argue that the synthesis of the concept of DBECO emerged in 2002 by adding “digital” in front of Moore’s \citep{moore1993predators} “business ecosystem”. Figure \ref{fig:TypesOfSECO} can be read as follows: a) each arrow comes from where the theoretical foundation of the SECO directions comes from, and to where the influence of the SECO theory goes; and b) the directions of the arrows implies that DECO, DBECO, BECO, and ECO influence SECO. It does not mean that SECO does not influence other kinds of SECO. DATAECO comes from a SMS conducted by \cite{Oliveira2019investigations} intended to function as a snapshot of the research in the field of Data Ecosystem, the authors state a DATAECO being another instance of a BECO, a DBECO, or a SECO. A DATAECO is a kind of ecosystem that shares network and co-evolution characteristics from the previous ecosystems; otherwise, it differs from previous ones because DATAECO does not rely on an explicit common platform in which different actors can collaborate. 

In fact, in SECO the products are software components or services. However, in DATAECO, the product is data. This is the justification for not including this study in the studies selection. Conversely, it brings to the SECO community the necessity to further investigate the results from others ones, mainly from the DECO and DBECO communities. Actually, it was already done by S1, S4, S8, S9, S10, S11 and S15 at including Management of Emergent Digital EcoSystems (MEDES) as source of their research strategy or selecting primary studies or references from there. This importance is emphasized by S7 that provides as limitation the fact or not including health of digital, mobile and business ecosystem in their search strategy. According to S7, it would broaden the picture of the whole ‘ecosystem health’ concept.  

The bottom of Figure \ref{fig:TypesOfSECO} exemplifies the kinds of SECO explicitly cited by the secondary studies. According to S9, Federated Embedded Systems Ecosystems (FESE) is an ecosystem where traditional embedded systems can be dynamically extended with plug-in software components, which can be provided by external developers. S11 argues that OSSECO is a growing research field in SE. Furthermore, it can be considered as a relevant topic of interest from the industrial perspective. OSSECO has many characteristics such as being placed in heterogeneous surroundings such as economic, social and technical environment whose keystone player is an OSS community around a set of projects in an open-common platform. S11 authors indicates that OSSECO is a specialization of SECO, BECO and DEBECO definitions. The same occurs with MSECO, that according to S6, is defined as a set of a collaborative systems, users and developers around a platform known as Mobile Applications Store or App Store. Besides, it inherits characteristics from natural ecosystems. S17 stands that MSECO are a specialized type of SECO having their focus in the mobile applications market. Another type of ecosystem found is the Technological health Ecosystem. S20 argues that health research initiatives have evolved from user-centered monolithic solutions into collaborative partnerships of different stakeholders that gather around different technological platforms. The authors states that these collaborative environments have evolved into what is known as SECO or ‘technological ecosystems’. As other kinds of ECOs, these ecosystems share some aspects of SECO such as a business-oriented point of view as a network of actors, organizations, and companies.

\subsubsection{RQ2.2: Which themes have been addressed by secondary studies on SECO?}

\begin{figure}[h]
	\centerline{\includegraphics[scale=0.45]{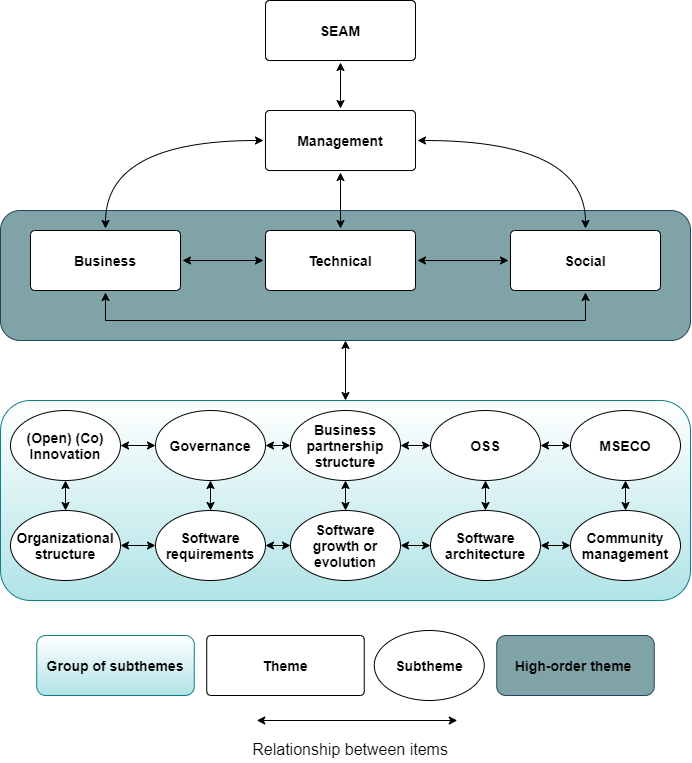}}
	\caption{SECO thematic map.}
	\label{fig:SECOThematicMap}
\end{figure}

In relation to thematic analysis, we applied an adaptation of the hierarchical structure of themes from the works of  \cite{cruzes2011b} and \cite{connelly2016}. Results were defined as one higher-order theme (Software Ecosystems), five themes and ten sub-themes. Themes and sub-themes, as well as the relationship between them can be seen in the thematic map presented in Figure \ref{fig:SECOThematicMap}. Furthermore, we defined dozens of categories according to each sub-theme and are presented in Table \ref{tab:SubthemesAndCategories}. The Figure \ref{fig:SECOThematicMap} illustrates the thematic map provided to SECO. In the center, we identified the high-order theme as being the SECO in the three-dimensional perspective which each dimension was mapped as a theme. Thus, even though each theme can be seen separately, each one is always interrelated with others. Through the results of the thematic analysis, we verified the need to identify management as a theme at the same level of business, technical and social. It is important to quote the distinction between both themes and dimensions. The identified themes should not be seen as new SECO dimensions. Rather, they are outcomes of the thematic synthesis process. In this sense, the management theme provides the connection between three-dimensional perspective views of SECO to a new one that was also identified by this research: Software Ecosystems Assessment Models (SEAM) theme. All themes are represented by rectangles. 

At the bottom part of image (Figure \ref{fig:SECOThematicMap}), represented in ellipses, there are the subthemes identified by the analysis of the secondary studies. The double arrows between items states that there is a bidirectional communication regarding the themes of the high-order theme to one or more subthemes.
This does not mean that the secondary studies explicitly identified all these relationships. Rather, we derived it as output of the thematic analysis. With respect to the subthemes, each one has relationships to another. Even though we recognize, for instance, that governance is more related to the business theme, there is no chance to not recognizing that the governance mechanisms influence both technical and social themes. As identified by S22, governance mechanisms provide implications to the interactions of the players of the SECO by using aspects of open innovation as an approach for managing the involved community in the SECO. Thus, for the sake of simplicity, we did not build a network map because there would be numerous connections between the subthemes and the themes and between the subthemes the categories. In addition, the interpretation of these connections could become very complex. This is the justification for not creating a thematic or a network map that would encompass all categories grouped for each subtheme.  

\begin{flushleft}
    \begin{longtable}{|c|l|l|}
    \caption{Subthemes and categories of the thematic analysis.} \label{tab:SubthemesAndCategories}\\
    \hline
    \multicolumn{1}{|c|}{\textbf{ID}} & \multicolumn{1}{c|}{\textbf{Subtheme}} & \multicolumn{1}{c|}{\textbf{Categories involved}} \\
    \hline
    \endfirsthead
     \caption{Subthemes and categories of the thematic analysis (Cont.)}\\
    \hline
    \multicolumn{1}{|c|}{\textbf{ID}} & \multicolumn{1}{c|}{\textbf{Subtheme}} & \multicolumn{1}{c|}{\textbf{Categories}} \\
    \hline
    \endhead
    %\hline
    %\multicolumn{3}{l}{Cont.}\\
    %\endfoot
    %\multicolumn{3}{l}{Fim.}\\]
    %\endlastfoot
    ST01 & \begin{tabular}[c]{@{}l@{}} (Open) (Co)\\ Innovation \end{tabular} & \begin{tabular}[c]{@{}l@{}} leveraging SECO members’ motivation; interactions among SECO members;\\ promoting interactions among SECO members; promoting cultural change; \\ strengthening the collaboration with academia; knowledge sharing; \\ promoting cultural change and contribution to something new;\\ attracting and maintaining new partners that can be individuals;\\ communities or even organizations; promoting the software platform;\\ innovating the software product/platform; opportunities for learning\end{tabular} \\ \hline
    ST02 & Governance &	\begin{tabular}[c]{@{}l@{}} analysis; openness; quality; architecture; health preservation; productivity; \\ community of contributors; attract and maintain new partners for SECO; \\ service choreographies in a service-oriented architecture (SOA); robustness; \\ quality of an open source software ecosystem (OSSECO); niche creation;\\ promotion of innovation; sharing knowledge; health improvement\\ verification and validation of services; and management of licenses \end{tabular} \\ \hline
    ST03 & \begin{tabular}[c]{@{}l@{}}	Business\\ partnership\\ structure \end{tabular} & value creation; proprietary software; and licensing management. \\ \hline
    ST04 & OSS & \begin{tabular}[c]{@{}l@{}} suitable architectures to support open source style development; reliability; \\ how to support the business strategy; open and flexible an architecture; \\ platform interface stability; evolution management; security; \\ risks for businesses at opening up the architecture to outsiders;\\ approaches for developing, acquiring, using and commercializing software\\ in a OSS context; symbiotic relationship between players;\\ modelling and analysis techniques; quality models; standard definitions;\\  promotion of sustainable development; health measurements;\\ health properties; and positive interaction between different organizations. \end{tabular} \\ \hline	 
    ST05 & MSECO & \begin{tabular}[c]{@{}l@{}} openness characteristic; OSSM; reusability issues; adaptability issues; users;\\ developers; community; evangelists; application store; platforms evaluation;\\ licensing policies; network level security; quality processes; software security;\\ application market or platform; commercialization of mobile devices;\\ contribution to something new; fun during the development process;\\ competition as an intellectual stimulus; learning and improvement skills;\\ gain reputation in the developers community; embedded software;\\ identification and commitment to the development community;\\ satisfaction of developers; knowledge exchange among developers;\\ developer self-management; advantages of using this mobile ecosystems;\\ size and quality of the developer community; and hackathons.\end{tabular} \\ \hline
    ST06 & \begin{tabular}[c]{@{}l@{}} Organizational\\ structure \end{tabular} & \begin{tabular}[c]{@{}l@{}} orchestration; decision making; technological infrastructure;\\ and keystone participation. \end{tabular} \\ \hline	
    ST07 & \begin{tabular}[c]{@{}l@{}} Software\\ requirements \end{tabular} & \begin{tabular}[c]{@{}l@{}} requirement elicitation; requirements verification; requirements validation;\\ coordination; communication; geographic distributed projects;\\ different stakeholders; quality assurance; system architectural design;\\ ecosystem management; stakeholder requirements definition;\\ mobile applications; quality assessment of OSSECO; quality of MSECO; \\ operationalizing quality requirements; establishing quality standards;\\ openness of entry requirements; identifying and analyzing requirements;\\ creating and managing software systems; non-functional requirements;\\ requirements engineering and quality aspects; analysis of emerging ideas;\\ identifying stakeholders and roles for requirements elicitation;\\ requirements communication or negotiation; conflict management;\\ conflict analysis; requirements prioritization; requirements selection;\\ performance;  testability; security; formalism; requirements testing. \end{tabular} \\ \hline	 
    ST08 & \begin{tabular}[c]{@{}l@{}} Software\\ growth or\\ evolution \end{tabular} & \begin{tabular}[c]{@{}l@{}}  platform evolution; product evolution; software evolution; co-evolution;\\ attractiveness for new players; key architectural challenges; innovation;\\ platform interface stability; systems evolution in an ecosystem;\\ evolution management; heterogeneity of software licenses;\\ risks of dependence; understanding of software component licenses;\\ open architecture systems; symbiotic relationships between OSS actors; \\ system integration; assessment approaches; synergetic evolution;\\ ecosystem entropy; ecosystem reciprocity; relationships;\\ components, infrastructures, and services from other companies;\\ ecosystem health; keystone participation in SECO;\\ growth and evolution of SECO over time; niche-players’ requirements;\\ ecosystem architecture quality attributes; platform orchestration;\\ analysis of OSS evolution; processes of innovations; hackathons;\\ evolution of the apps or the central platform; measures;\\ governance strategies and managerial decisions taken by keystones. \end{tabular} \\ \hline	
    ST09 & \begin{tabular}[c]{@{}l@{}} Software\\ architecture \end{tabular} & \begin{tabular}[c]{@{}l@{}} maintenance properties; code visualization; evolution of the ecosystem;\\ opening the platform architecture; keystone participation in SECO;\\ quality of a system being built; systems, system components, and actors;\\ competitive architectures for innovation; health; health threats;  defects;\\ platform architecture; quality attributes; technological architecture;\\ architectural analysis of the ecosystem;  open platform architecture;\\ business and management relationships; open architecture software system;\\ evolutionary paths of the system and architecture; ownership of the code;\\ interactions among all actors within the ecosystem; delivery mechanisms;\\ common infrastructure; cloud; tools for the development of new application;\\ managing the evolution of the enterprise architecture; open-innovation;\\ security mechanisms; connecting relationships between ecosystem actors. \end{tabular} \\ \hline
    ST10 & \begin{tabular}[c]{@{}l@{}} Community\\ management \end{tabular} & \begin{tabular}[c]{@{}l@{}} open-source community; community building; activities of coordination;\\ alliance forming; management of OSS communities; health of a SECO;\\ quality measures; positive relationships (symbiosis) between actors;\\ management processes; keystone participation in SECO;\\ correlation between the community and the keystone; hackathons;\\ maintenance capacity, sustainability and process maturity;\\  structures of collaboration; strengthening the collaboration with academia;\\ attracting new partners; knowledge sharing and opportunities for learning;\\ interactions among participants; promoting the software platform;\\ attractiveness for members of the community; sustainability of the SECO;\\ relationship between communities and the keystone of the SECO;\\ cultural change; forms or mechanisms to open the platform or the SECO;\\ assessing the software platform; leveraging participants’ motivation;  \\ the form of autonomy given for the SECO community; software projects;\\  community building of users and experts; open technological platform;\\ innovating the software product/platform; participatory contribution; \end{tabular} \\ \hline
\end{longtable}
\end{flushleft}

\begin{itemize}
    \item \textbf{ST1: (Open) (Co) innovation}
\end{itemize}

Open or co-innovation is the first subtheme correlated with the SECO in the three-dimensional perspective high-order theme. S5 points out that an aspect of innovation in ecosystems is a business innovation, S5 authors argue the need to research the best current practices for business innovation in the software domain. In this regard, S22 focused precisely of one of the research agenda items from S5. It focused on the benefits of hackathons for SECO. Hackathons can be considered an approach of open innovation that promotes several implications to the social and technical dimensions of SECO. From the social perspective, S22 argues that hackathons promote repercussions such as motivation, strengthening collaboration with academia, promoting interactions among participants, promoting cultural change, and knowledge sharing. Conversely, S16 focused on partnership models related to SECO. This study did not mention directly innovation as a partnership model, it nevertheless mapped “value co-creation” as a goal category of partnership models. Co-creation, in this regard, can be seen as a result of the use of hackathons, because it can be considered a form of governance mechanism (S15) for sharing knowledge, and promote innovation by attracting and maintaining new partners. They can be individuals, groups of people, communities or even organizations. Moreover, if the ecosystem behind the platform is OSS, the use of hackathons relates to development best practices (S11), innovating the software product/platform and promoting the software platform (S22). Thus, as identified by S1, SECO approach fosters the success of software, the co-evolution and innovation inside the organization, increase attractiveness for new players.

\begin{itemize}
    \item \textbf{ST2: Governance}
\end{itemize}

Governance is the second subtheme correlated with the SECO in the three-dimensional perspective. S1 was the first secondary study to map this concept in SECO secondary studies. The study highlighted some challenges by recommending some future research directions for SECO specifically related to OSS, governance, analysis, openness, quality, and architecture of ecosystems.  According to S1, the main question for SECO is to identify which are the best survival strategies in any ecosystem, regardless of the role that an actor plays in the ecosystem. This fact is not necessarily associated with a specific type of ecosystem or the business partnership structure adopted (an open source developer, an open source consortium, or the largest software vendor in the world), this fact is related to which approaches a keystone should act in the process of directing the ecosystem members.  Regarding governance aspects in SECO, this study identified that S4, S6, S07, S11, S12, S15 used the model developed by Jansen and Cusumano (2013) to identify some ecosystem characteristics such as health aspects, robustness, niche creation, and productivity. Furthermore, they provided, for each characteristic, a set of governance tools. S4 did not perform research about governance. Conversely, it focused on measuring the quality of an OSSECO. However, it is connected to the governance subtheme as identified by S7. S7 focused on SECO health. Nevertheless, in its findings, what can be observed it is a considerable relationship between the concepts of governance and ecosystem health as follows: “Ecosystem governance is argued to impact ecosystem health”; “Ecosystem governance leads to better ecosystem performance and health.”

In S9, authors presented the governance term related to verification and validation of services and service choreography in a service-oriented architecture (SOA). S11 emphasizes the concepts behind OSS and also highlighted the difference about the diverse methods of governance between ecosystems, when they are open or proprietary. When proprietary, it is usually led by the keystone. On the other side, when is an OSSECO, the community of contributors determines which contributions are accepted into the source code base.  S15 used thematic analysis and identified the relevant codes from primary studies, then, they merged the codes into key themes. They considered, for illustration, that governance mechanisms relate to other terms such as management, govern, control, strategy, orchestration and critical factors. Equally important are the connections that can be found in either the business theme or the technical. S15 indicates some governance mechanisms such as the management of licenses, sharing knowledge, and promotion of innovation. In this regard, in terms of licenses, S12 discussed that requirements engineering activity in SECO involves an understanding of software component licenses that are part of the composed system after the system integration. Thus, the software licenses aid and constrain the SECO evolution. Requirements engineering and licenses can be considered in the technical perspective of ecosystem. Besides, it is also in the business perspective because it is related to the evolution of a SECO. Governance can also be related to hackathons (S12), as they can be considered  a way to promote innovation, attract, and maintain new partners. As identified by S12, hackathons provide several benefits related to social dimensions do ecosystems. As a result, we consider governance sub-theme to be primarily focused to business theme, secondly focused to technical theme, and finally, with implications to social theme.

\begin{itemize}
    \item \textbf{ST3: Business partnership structure}
\end{itemize}

Regarding the type of product development ecosystem or platform, the literature has classified it in different ways. One of them has to do with whether the ecosystem, the platform or the community used the OSS model, closed/proprietary model or hybrid characteristic. S2 identified two approaches, FOSS ecosystems. S2 used the term SECO kind or SECO type to refer to the model of operation of the SECO. In S8, the author has employed the concept of business structure to describe these characteristics in relation to business by defining the means of value creation in the ecosystem. Related to the concepts, these studies used three concepts: proprietary: in which value creation in the ecosystem is based on proprietary contributions; open source: in which contributions are typically open to the rest of the actors or public; and hybrid: in which the ecosystem that supports both proprietary and open source contributions. To define the same concept, S16 calls it as the domain of the partnership model used by the ecosystem: “Open source”, “Proprietary”, and “Open Source and Proprietary”. S13 has used the term dimension to describe the same background. The authors consider two dimensions. The first one is in relation to license (open source and proprietary), and the second is related to kind of device that the software is used (embedded and mobile). S17 used the concepts brought by S8. From the perspective of this tertiary study and regarding the analysis thematic, this study proposes a definition of this subject by integrating the two meanings used by S8 and S16 to suggest the “business partnership structure” subtheme. It is because we recognize that the form of the license used in ecosystem has to do with what types of partnerships the keystone or the community will make and it is a business decision. Thus, it is under the business dimension.

\begin{itemize}
    \item \textbf{ST3: Business partnership structure}
\end{itemize}

4.2.2.4 ST4: Open source software (OSS)
As stated above, OSS is fully related to the business partnership structure subtheme. S1 has mapped it as one of the most common areas in SECO. This study argues that one characteristic of SECO is its linkage to Open Source Development Model (OSSM). According to S1, the literature in OSS reinforces the importance of collaboration among players to build a mature open source platform. S2 points out that OSSM is related to business and management aspects of SECO once there is an organizational and management entity that is responsible for monitoring, operational and decision-making part of the SECO whether it being a proprietary company, an open source community or a hybrid of the two. S5, in turn, mapped in the literature that risks are highlighted for businesses opening up to outsiders, third parties, or open source communities, but also benefits from doing so. It emphasizes the business partnership structure discussed by S8 and S16. From the aspects of mobile ecosystems, S6 provided an extension of S1 and mapped that OSSM is a characteristic of MSECO. It demonstrates that regardless of the type of platform behind the ecosystem, the OSSM is fundamental for the studies of ecosystems. Regarding the health of SECO, S2 provided a categorization about ecosystem health literature, and OSS is one of the four categories proposed. Jansen (2014) also studied health in the context of OSSM. He proposes an operationalization of OSSECO health measurements on the network level and project level, with several kinds of measurements. As observed in Table 7, health is topic in at least four secondary studies as follows: S7, S9, S10, and S14.

In S10, “open source ecosystem” is a term that is used to define the research strategy used by the reviews. S10 observed that the evaluation scenarios regarding health are composed mainly by OSSECO. This emphasizes the importance of studying OSSECO when studying their health properties. S11 is a mapping review especially devoted to the study of OSSECO. They state that OSS and SECO are two consolidated research field in SE. OSS influences the way organizations develop, acquire, use and commercialize software. Thus, SECO appear as a valid instrument to analyze OSS systems. S11 concluded that regarding the definitions related to OSSECO, there are five major concepts (i.e., ECO, BECO, DBECO, SECO, and OSSECO). In addition, they found that there are only three definitions of OSSECO, and proposed a new definition integrating the different definitions related. They state that an OSSECO is a SECO placed in a heterogeneous environment, whose boundary is a set of niche players and whose keystone player is an OSS community around a set of projects in an open-common platform. Among the elements that belong to an OSSECO, they identified that project, community, and source code are the most used. Furthermore, they sketched a taxonomy with three categories (i.e., OSS community, ecosystem network, and software platform) to classify the OSSECO terms. They also identified 27 instances of OSSECO that appear in their mapping. Among them, Eclipse and GNOME are the most frequently used. Despite the results, they conclude that existing research on several topics related to OSSECO is still scarce (e.g., modelling and analysis techniques, quality models, standard definitions, etc.).

This study identified that OSSM has several implications for the platform owners, their contributors and users. For instance, S12 argues that Microsoft made the PowerShell tool as an open source product to keep developers interested in the Windows platform. As comparison, Google released Cloud Tools for PowerShell to make Google's cloud more attractive to .NET developers. It can be seen as a positive symbiotic relationship between players. Symbiosis can be defined as the positive interaction between different organizations, users or platforms. In this context, OSSECO provides a symbiotic interaction that keeps the participants of the SECO living in a way that one depends on and fosters the life of the other. This concept is also evidenced in business literature of ecosystems. For instance, Moore (1993) argues that the scope of a Business Ecosystem is the set of positive relationships (symbiosis) between actors who work together around a core technology platform. It can be seen as a strategic concern by following the idea stated by S15 that states that orchestrator (or keystones) that can balance their own interests by bringing joint benefits for other players is likely to create healthy ecosystems. The platform owner must promote the sustainable development of the ecosystem by defining strategies and orchestrating the activities of players. In this context, governance appears a crucial aspect for open source communities as proprietary.

\begin{itemize}
    \item \textbf{ST5: Mobile software ecosystems (MSECO)}
\end{itemize}

S1 mapped several research directions in the literature that includes aspects related to architecture, social networks, modeling, business, mobile platforms and organizational-based management. In reference to mobile, Mobile Learning Ecosystems (MLES) and Mobile Ecosystem, S1 figured these terms in the list of the main terms related to the SECO field. In connection to mobile, S1 also identified the OSS and the openness characteristic of a SECO that relates to how openness in software affects and influences the success of business. In this context, S2 also identified these two aspects. First, an example that addresses reusability and adaptability issues in MLES. The second is the one that analyses the mobile platforms for evaluation of their level of openness taking into consideration their licensing policies. In terms of requirements engineering and quality processes, according to S12 mobile ecosystems might have different quality focus than embedded SECO. While in embedded SECO, the focus is on network level security, in mobile SECO the focus is on software security. In connection with the kinds of device in which the software is released, S13 discusses about the domain of the studies focus at, the license and device. They also argued differences between the mobile device and others specific embedded software ecosystems. S17 is the first secondary study dedicated to MSECO Literature. S17 points MSECO as a specialized type of SECO, and have their emphasis in the mobile application market or platform. In reference to the growing of MSECO, it is related to the increasing commercialization of mobile devices. 

As elements or actors of a MSECO, the authors identified the following stakeholders: users, developers, community, and evangelist. They have symbiotic relations with the platform (technological architecture on which the MSECO operates), the platform market (online store where users can search and acquire applications), and the applications in the mobile market. S17 focused to classify its list of primary studies by using the SECO in the three-dimensional perspective. The results indicate that studies in MSECO have used hybrid perspectives regarding the dimensions such as technical and business alone as far as social technical \& business, technical \& social, and business \& social. S17 concludes that although MSECO are extremely relevant today, little is known about the field of research in MSECO. In addition, S17 addressed some opportunities for future directions to fill in the identified gaps, mainly those related to the social dimension of SECO in the context of MSECO. S19 seems to have heard the predecessor, S17. Although it did not mention the precursor directly, it exactly focused on an aspect aimed by S17 as further studies: social aspects of MSECO. In this study, 20 developers were interviewed in order to understand how social aspects influence MSECO developers. Both S17 and S19 theorizes at the same background. Thus, these studies used the same elements or actors of a MSECO, as namely: platform (technical aspect), application store (business aspect), applications (technical or business aspect), developers (social aspect), users (social aspect), community (social aspect), and evangelists (social aspect). They identified several social factors, as namely: contribution to something new; fun during the development process; competition as an intellectual stimulus; learning and improvement skills; gain reputation in the developer community; identification and commitment to the development community; satisfaction of developers with their MSECO; knowledge exchange among developers; developer self-management; advantages of using this MSECO; and size and quality of the developer community.

\begin{itemize}
    \item \textbf{ST6: Organizational structure}
\end{itemize}

This subtheme of SECO is derived from S8 that used the work of Christensen et al. (2014) to provide the SECO classification. S8 discussed SECO characteristics by identifying which ecosystem’s organizational structure governs the ecosystem.  According to S8, one relevant aspect of ecosystem orchestration is the decision making, i.e., who and how makes decisions in software ecosystems. In this context, S8 proposes a set of archetypes for orchestration decision making, as namely: monarchy, federal, collective, and anarchy. The first archetype is the one in which only one actor orchestrates the ecosystem. Usually, this actor has a boundless impact on the shared technological infrastructure provided to the SECO. The second archetype is known as federal and the orchestration is led by a set of representative stakeholders. The third archetype, collective, it is when the ecosystem is orchestrated through processes concerning all the actors; and finally, the last archetype, anarchy, is when the ecosystem is characterized by the lack of a general orchestration and each actor acts on their own, based on local needs. To do so, S8 was inspired by the theme of IT governance to provide an ecosystem classification with respect to the approach that the keystone acts in the ecosystem. Although this characterization is not fully adopted for many other secondary studies, we identified it as an important characteristic once the keystone plays an important role in the SECO environment. S19 is a study that recognized the organizational structure of technological ecosystem (a synonymous used for SECO) when it states that this concept is related to how the interaction and organization of actors and software are governed, which is in turn related and dependent on the ecosystem.

\begin{itemize}
    \item \textbf{ST7: Software requirements}
\end{itemize}

Requirements engineering stand as a challenging aspect of SECO as stated by S1 when it identified that there are technical and socio-organizational barriers for coordination and communication of requirements in geographic distributed projects. This is also identified by S2 that argues that requirement elicitation appears as an interesting challenge in the SECO concept as the stakeholders are multiple and distant from the central ecosystem management. These challenges are related to communication among the different stakeholders because according to S2, how knowledge is transferred to the different roles of a SECO is, inclusively, a topic of research. Challenges are also mapped by S11 regarding quality assurance in SECO (e.g., stakeholder requirements definition and system architectural design). Requirements engineering also relates to compliance with the business goals. S4 argues that the quality assessment of OSS ecosystems is of vital importance because quality assurance is a way to prevent bad decisions and avoid problems. In this context but in the field of mobile, S6 identified relationships between quality of MSECO that can be applied to the context of mobile applications to prevent bad decisions, solve problems and to allow verifying requirements and business objectives. S11 also mapped quality and requirements engineering in the same context when it discusses OSSECO monitoring. A solution for monitoring OSSECO is to link the gathered values with adopter needs by operationalizing quality requirements of open source solutions. 

With reference to the approaches of governing a SECO, requirements engineering also plays an important role. S15 discusses that orchestrator players in SECO can control by defining entry requirements, establishing quality standards, and through certification. Thus, the procedure that a keystone operates the openness of entry requirements can be seen as mechanisms to govern SECO. With respect to social dimension, S1 also mapped the correlation between requirements engineering and this dimension. According to them, once SECO must satisfy diverse requirements of users, identifying and analyzing requirements, creating and managing SECO can be considered a social activity. S12 is a review that devoted its execution to fully research about requirements engineering. The study presented a mapping study on the issues of requirements engineering and quality aspects in SECO and analyzed emerging ideas. S12 outcomes indicate that the most addressed topics on requirements engineering in SECO are identifying stakeholders and roles for requirements elicitation and requirements analysis. In addition, S12 findings indicate that the most addressed topics on nonfunctional requirements are related to security, performance and testability. S12 mapped several research activities across requirements engineering activities such as elicitation, analysis, specification, validation, and management. They provide a mapping between activities and topics in the context of requirements engineering in SECO. For elicitation, the following topics were mapped: goal modeling, reference model, Nonfunctional requirements, Identifying, stakeholders’ roles, identifying relationship, and policies. For analysis, requirements communication or negotiation, conflict management, conflict analysis, requirements prioritization, and requirements selection. For specification¸ notation semantics and modeling approaches. For validation, model formalism, requirements verification, validation and testing. Moreover, for management, global requirements engineering and management practices.

\begin{itemize}
    \item \textbf{ST8: Software growth or evolution}
\end{itemize}

This theme is with respect to the following aspects to SECO: software, product and the platform evolution. A large amount of research has set evolution as one of the core aspects of SECO. According to S1, SECO perspective fosters the success of software. In this context, the co-evolution and innovation inside the organization, increase attractiveness for new players. Nervetheless, there are several key architectural challenges such as platform interface stability, and evolution management. For instance, the heterogeneity of software licenses and system evolution in an ecosystem is identified as a challenge of SECO perspective once organizations must manage these issues to decrease risks of dependence. S2 also identifies it by concerning to the issue of software licensing in open architecture systems. S2 recognizes that changes in licenses on different versions of the same component or in the evolution of a software system and propose a structure for modeling software licenses. Software licenses are also subject of discussion by S12. It argues that the requirements engineering activity in SECO involves an understanding of software component licenses that are part of the composed system after the system integration. Moreover, the software licenses of the components both aid and constrain the SECO evolution. Concerning assessment approaches, the concept of synergetic evolution by Li et al. (2013) and mapped by S4 comes as the ability of the subsystems that constitute the whole ecosystem to form a dynamic and stable space-time structure. S4 states that measures such as ecosystem entropy, distribution over the species, and ecosystem reciprocity can be used to evaluate synergetic evolution. S6 states that some relationships appear with software evolution and start to comprise components, infrastructures, and services from other companies. In this context, S7 emphasizes that characteristics of ecosystem health include both growth and evolution over time. 

Regarding the importance of the keystone, S9 argues that this player might influence the evolution of the platform leading to more advantages for itself but disadvantages for the smaller niche players as well as for the ecosystem as a whole. Furthermore, evolution of the platform is a large extent connected to understanding the niche-players’ requirements and balancing these against ecosystem architecture quality attributes. This is also emphasized by S22 that argues that keystone is the actors that governs the evolution of the ecosystem by defining rules of access to the platform and orchestrating the creation of new solutions. With respect to innovation, S22 discusses that Tech companies from all sizes have increasingly integrated hackathons into their software development work to support the product test and evolution. In this scenario, a hackathon can be seen as a strategy to support ecosystem evolution once they are developed to create platforms so that third parties can develop new software solutions and in doing so extend the current product portfolio.
With respect relationships of players, S11 states that OSS projects typically provide access to several kinds of data sources to extract information about their evolution and the symbiotic relationships between OSS actors. Thus, this results that OSS projects classically afford community accessibility of historical data, which simplify the process of analysis of OSS evolution. Still with regard to relationships of actors, it is important to mention the importance of the processes of innovations such as hackathons that according to S22, they can foster longer collaborations after the event. Thus, smaller companies can perceive hackathons as a chance to join an ecosystem by interacting with bigger players and consequently providing the evolution of the apps or the central platform. Analysis of the results in secondary studies highlights the importance of the better understanding of evolution aspects of the SECO. This is evidenced, for instance, by S15 that discusses governance strategies and managerial decisions taken by keystones can affect the healthy evolution of the entire ecosystem. Thus, the use of measures can provide operational indicators on how software ecosystems are evolved and governed.

\begin{itemize}
    \item \textbf{ST9: Software architecture}
\end{itemize}

According to S1, software architecture is one of the eight most common areas studied in SECO. It is connected to the technical dimension once it explores decision-making aspects and architectural property maintenance, for example, using design and code visualization. Once technical dimension is associated with planning the process of opening the platform architecture, software architecture has several implications to the business perspective or dimension. Concerning the single systems, S2 argues that software architecture is seen as essential in determining the quality of a system being built. S2 provides their own definition of software ecosystem architecture as being the structure or structures of the software ecosystem in terms of elements, the properties of these elements, and the relationships among these elements. It recognizes elements such as systems, system components, and actors. S5 comes with a research agenda regarding competitive architectures for innovation. It argues that an important aspect of the innovation system is competitive architecture. It discusses that it is important to understand which parts of the organizational capability can enable open innovation in a SECO. Thus, there is a need to further understand the competitive architecture and organizational capabilities that foster open-innovation. 

As several other aspects, themes, and subthemes in this study, architecture are also related to the health concept of the SECO. In this context, S7 identified that an ecosystem architecture can pose risks that endanger the health of the entire ecosystem; an architectural analysis of the ecosystem can reveal health threats. As health is closed to the concept of quality, S9 states that an ecosystem architecture forms the basis of the ecosystem platform and infrastructure and the quality attributes of the architecture are thus of high importance. In relation to platform, S9 argues that there is a need for protective mechanisms in the platform architecture to deal with differentiated access to resources for extensions depending on niche player privileges, to isolate extension execution and to support data integrity. In addition, evolution of the platform is largely connected to understanding the niche-players’ requirements and balancing these against ecosystem architecture quality attributes. S8 identified the most common remaining keywords for the group ‘SE’ is “architecture” along with the similar “architectural” and “architectures.” S8 classified the primary studies in accordance with the software ecosystem architecture groups, as namely: SE, business and management, and relationships. Furthermore, S8 used the work of Christensen et al. (2014) to expand the concept of ‘software ecosystem architecture’ and used it to propose the modeling of software ecosystems.

Software architecture plays an important role with respect to OSS. In this context, S12 points out that an open architecture software system can provide niche developers with guidance for identifying evolutionary paths of the system and architecture. The same occurs with respect to security. S12 identifies that a good architecture can minimize the likelihood for malicious code to affect the whole system, but the defects can only be minimized to the extent that the security mechanisms are correctly implemented. S15 also identifies the connection to OSS. According to the authors, without an open platform architecture, extenders cannot extend it. They will always exist concepts such as ownership of the code and supporting tools to use and evolve the ecosystem. From the viewpoint of governance, S15 linked the keystone actor to the processes of managing the evolution of the enterprise architecture and the interactions among all actors within the ecosystem. In this context, S15 also identified a connection to the social dimension once there is a need for further study of enterprise architecture and delivery mechanisms that enables software ecosystems. According to S15, keystones must understand developers’ motivations and expectations to adopt appropriate governance mechanisms.

Software architecture appears relevant to any kind of ecosystem. S17, for instance, states that the concept of platform refers to the technological architecture on which the MSECO operates. This architecture encloses communication protocols and resources available to players. Thus, without the platform, the ecosystem mobile would not work neither evolve. An aspect of the software architecture is the finding of S19 that identified that the main ecosystem architecture types considered in the literature consist of ecosystems that share a common infrastructure such as cloud or mobile. In these infrastructures, the technology is shared with all other players. Thus, the keystone delivers tools for developing new applications while maintaining freedom on the contributions. Despite the importance of the software architecture, this study identified that there is a lack of concrete classifications in the literature in terms of technological ecosystems’ architecture no matter what their scope. However, S19 authors recognize three main structures could be found in software ecosystems: common software or technological platform (technical); business or interests (business), and connecting relationships between ecosystem actors (social).

\begin{itemize}
    \item \textbf{ST10: Community management}
\end{itemize}

Lungu (2009) argues that SECO is a collection of “software projects” which are developed and co-evolve together in the same environment. This environment can be physical, like in the case of a company or a research group that has a geospatial identity, but can also be virtual, like the projects that are part of an open-source community. In this respect, S1 argues that the management of these communities is crucial once it is related to the sustainability of the SECO. These results indicate, for instance, that community building, alliance forming, and participatory contribution are essential activities that enable OSS projects to persist without central corporate authority. This is strongly related to the social dimension of SECO. In that regard, according to S2, an open technological platform in combination with a set of management processes and business models cannot create a SECO without the social aspect. S7 also identified the correlation between the community, the keystone, and the sustainability of the SECO. Ecosystem sustainability is defined as keystone activities to maintain the community. In S4, on the QuESo model, the community is a dimension related to the quality characteristics provided. S4 identified three main communities’ characteristics, namely, maintenance capacity, sustainability and process maturity. In addition, they provided several measure definitions centralized on the SECO community. As example, there are such as the number of inquiries or feature requests (number of inquire or feedbacks received for the oss community); community effort (the combined effort of all members belonging to community); date of last commit (date of last commit of a project/community), and geographical distribution (geographical distribution of community members).

S6 puts community as a formal element of MSECO. It argues that communities can be as the set of users, set of developers and set of experts. All of them can be as the structure of collaboration and activity coordination within an ecosystem, composed of internal and external actors. S7 points out the same direction when one of their findings states that a community (of developers, experts and users) is vital for the health of a SECO and that a keystone player’s mission is to promote the overall health of an ecosystem. In this regard, S10 highlights a similar concept when it uses the work of Iansiti and Levien (2004a and 2004b) to discuss health and communities. In this sense, health is the ability to grow the ecosystem and keep it attractive for all members of the community. Communities seem present in any type of ecosystem as stated by S11 that also puts community as one of the common elements of SECO, in special, OSSECO. They argue a community as a set of actors (i.e., complex organisms in ECO, business world organisms in BECO and DBECO and collections of products, projects, software solutions, and businesses in SECO and OSSECO). For instance, S15 argues that a fundamental governance decision that orchestrates must make is how much power is given to the community and how much control it keeps for itself. Moreover, the S15 still-discussed mechanisms to open the platform or the SECO. For instance, how autonomy will be given for the SECO community to make their own decisions independently? These autonomies can be notorious, once, according to S17, in some OSSECO the decisions are taken by the community and no company has control over its business model.

With respect to business and social mechanisms. S22 identified several aspects that corporate hackathons can benefit SECO, as namely: hackathons can attract new partners; they entail ecosystem growth by building a community of users and experts; hackathons promote knowledge sharing and opportunities for learning; hackathons promote interactions among participants; promotes cultural change; leveraging participants’ motivation; strengthening the collaboration with academia. In addition to the business and social benefits, they mapped other technical implications to the use of hackathons, such as innovating the software product/platform; promoting the software platform; and assessing the software platform. All these benefits are made possible with the active management of the ecosystem community and with the strong keystone participation. These results emphasize the importance of the communities and highlight the connection to the business ecosystem paradigm that SECO is based on. For instance, Moore (1993) argue that the scope of a Business Ecosystem is the set of positive relationships (symbiosis) between actors who work together around a core technology platform. Moore (1993) complements that all actors in a Business Ecosystem are connected and share the success or failure of the network as a whole. These relationships (symbiosis) between actors and this network are compound of communities.

\subsubsection{RQ2.3: What are the gaps of research topics pointed by secondary studies on SECO?}

Research topics discussed in secondary studies cover a wide variety of topics related to SECO. However, we identified several gaps for the field were according to the outcomes of this study. Table \ref{tab:SECOGaps} summarizes the main gaps according to the secondary studies and their research topics.

\begin{center}
    \begin{longtable}{|l|l|l|}
    \caption{SECO gaps.}\label{tab:SECOGaps}\\
    \hline
    \multicolumn{1}{|c|}{\textbf{Topic}} & \multicolumn{1}{c|}{\textbf{Gap}} & \multicolumn{1}{c|}{\textbf{\begin{tabular}[c]{@{}l@{}} Categories\\ involved \end{tabular}}} \\ \hline
    \endfirsthead
     \caption{SECO gaps. (Cont.)} \\ \hline
    \multicolumn{1}{|c|}{\textbf{Topic}} & \multicolumn{1}{c|}{\textbf{Gap}} & \multicolumn{1}{c|}{\textbf{\begin{tabular}[c]{@{}l@{}} Categories\\ involved \end{tabular}}} \\
    \hline
    \endhead
    %\hline
    %\multicolumn{3}{l}{Cont.}\\
    %\endfoot
    %\multicolumn{3}{l}{Fim.}\\]
    %\endlastfoot
    General	& Theories, methods and tools specific to SECO problems & [S1] [S8] [S11] [S19] \\ \hline	
    General	& SECO Modeling & [S1] [S11] [S19] \\ \hline
    General	& Social aspects of SECO & [S1] [S9] [S11] \\ \hline
    General	& Opening of SECO & [S1] \\ \hline	
    General	& Relationship of SECO with other ecosystems & [S2] \\ \hline	
    General	& Intellectual property and licensing in SECO & [S2] \\ \hline	
    \begin{tabular}[c]{@{}l@{}} Software\\ Architecture \end{tabular} &	Software architecture adequate for SECO & [S1] [S9] [S13] [S19] \\ \hline \begin{tabular}[c]{@{}l@{}} Software\\ Architecture \end{tabular} &	Quality attributes for SECO architecture & [S9] \\ \hline	
    OSS & OSSECO Characterization and Modeling & [S1] [S11] \\ \hline	
    OSS & Quality in OSSECO & [S11] \\ \hline	
    measures & A broader understanding of Key Performance Indicators (KPI) & [S3] \\ \hline	measures & Identification, analysis and evaluation of KPI & [S3] \\ \hline	
    \begin{tabular}[c]{@{}l@{}} Open\\ Innovation \end{tabular} & Primary innovation in SECO & [S5] \\ \hline	
    \begin{tabular}[c]{@{}l@{}} Open\\ Innovation \end{tabular} & Competitive architectures for innovation in SECO & [S5] \\ \hline	
    \begin{tabular}[c]{@{}l@{}} Open\\ Innovation \end{tabular} & Procedures and techniques that support innovation in SECO & [S5] \\ \hline	
    \begin{tabular}[c]{@{}l@{}} Open\\ Innovation \end{tabular} & \begin{tabular}[c]{@{}l@{}} Software innovation and collaboration across organizational\\ frontiers \end{tabular} & [S5] \\ \hline	
    \begin{tabular}[c]{@{}l@{}} Open\\ Innovation \end{tabular} & \begin{tabular}[c]{@{}l@{}} SECO processes and practices that support business\\ innovation and Software innovation combined \end{tabular} & [S5] \\ \hline	
    Mobile & Communities and evangelists in MSECO & [S17] \\ \hline	
    Mobile & Deeper understanding and classification of MSECO elements & [S17] \\ \hline	
    Mobile & Social Aspects in MSECO & [S17] \\ \hline	
    Mobile & Motivation of users in MSECO &	[S18] \\ \hline	
    Health & Analysis, measurement and evaluation of the health of SECO & [S2] [S7] [S10] \\ \hline	
    Health & Definition of the concept of health in SECO & [S7] [S10] \\ \hline	
    Health & \begin{tabular}[c]{@{}l@{}} Appropriate practices, measures and instruments to conduct the\\ SECO health assessment process \end{tabular} & [S10] \\ \hline	
    Quality & Quality Assurance in SECO & [S1] [S2] \\ \hline	Quality & Quality analysis according to the SECO domain & [S12] \\ \hline	
    RE & \begin{tabular}[c]{@{}l@{}} Coordination, communication, validation and management of\\ requirements in SECO \end{tabular} & [S1] [S11] [S12] \\ \hline	
    RE &  Definition, selection and prioritization of Requirements in SECO & [S2] [S9]  \\ \hline	
    RE & Tools to model and analyze requirements in SECO & [S12]  \\ \hline	
    RE & Standards to specify requirements in SECO & [S12]  \\ \hline	
    RE & Guidelines for the process-engineering requirements in SECO & [S12] \\ \hline	
    Governance & Understanding governance in SECO &	[S1] \\ \hline	
    Governance & Governance to foster innovation in SECO & [S2] \\ \hline	
    Governance & Common Vocabulary in SECO Governance & [S15] \\ \hline	
    Governance & Practical Guidelines for SECO Governance & [S15] \\ \hline	
    Governance & \begin{tabular}[c]{@{}l@{}} Analysis on the interaction between governance\\ mechanisms and health measures in SECO  \end{tabular} & [S15] \\ \hline	Governance & Governance in developer ecosystems & [S15] \\ \hline	
    Governance & Governance in OSSECO & [S15] \\ \hline	
    Governance & Ecosystem interactions & [S15] \\ \hline
\end{longtable}
\end{center}

Table \ref{tab:SECOGaps} presents the principals gaps that have been pointed in secondary studies related to SECO according to how each research topic presented in Figure \ref{fig:TopicOfResearchOverTheYears}. The lack of theories, methods and specific tools for problems of SECO were quoted by S1 that highlight the necessity of development of tools and specific strategies to software analysis to SECO. S8 highlight the enterprise to development of these approaches, however, affirms that the solution, normally, is highly coupled to an ecosystem and is not easily transferred to other. S11 highlight the lack of established theoretical basis to SECO and quote sociotechnical as an opportunity of research to help to understand OSSECO. For S12 exist the necessity to define clear and formal directions to develop technological ecosystems and the necessity to develop tools that evaluate the performance of these ecosystems.

SECO modeling received a highlight on quotations in studies and it was approached as a lacuna in S1, which focus in the necessity of a universally set accepted of modeling methods for SECO. S11 quotation the challenge rose by S1 and confirms its importance to OSSECO, focusing the importance of modeling great network related to evaluation and dynamic of ecosystems, thus, one of main challenge inside the domain of SECO. For S19, different paradigms still need to use to capture a view more realistic, filling the gap between for the model of ecosystem and ecosystem that need to be modeled. S11 even quotes the lack of specific techniques of modeling to SECO.

Three studies (S1, S9, and S11) mentioned gaps related to social aspects of SECO. S1 described that it is necessary a correct engagement of keystone in the social dimension and, that the infrastructure development and tools for social interaction is an opportunity of research for the area. S9 shows the necessity to implement a social support to develop of the central platform. S11 highlights that information about the social behaves in ecosystem must be registered and that techniques as machine knowledge and text mining can be used to identify social issues in OSSECO. Still related to SECO general aspects, it has been indicated the necessity of additional researches about how software opening affect and influence the business success (S1); the relations of SECO with other ecosystems (S2); and the intellectual property and license of SECO (S2).

SECO architecture was also a gap pointed by secondary studies. S1 has brought questioning about how software architecture can be projected to attend SECO, because need to be open, adaptable, functional, evolutional, easy maintenance and embrace several domains. S9 brings an embracing list of challenges of specific research to ensure the quality in SECO. For S9, the challenge is related to interoperability of the interface provided for the central platform to enable an independent work of multiple agents. Besides, for the agents, the attributes of architecture quality are of great importance. On this manner, additional researches are necessary for a better understanding in which quality attributes are important to provide a platform that support a quality ensures inside SECO. S13 affirms that there is not a systematic review focused in software architecture and oriented engineer for models in SECO. S19 highlight that is necessary more works that treating about an open architecture for technological ecosystems. 

Related to OSS, S1 affirms that several questions can be answered through OSSECO study, such as governance and SECO analysis, for this reason, more research should be done. S11 highlight the lack of a method set of modeling universally accepted for SECO and that are challenges for this area: the great network modeling, development of visualization tools of scalable models and a study of evolution and dynamic of SECO. S11 still does an analysis when an immaturity of quality management and operationalization of SECO. For the authors, OSSECO quality is great different of quality traditional norms and suggest that new researches can analyze if OSS quality models, that suggested, due to inability of traditional quality models, can be base of OSSECO quality models.

S3 suggest that wider comprehension of KPI would increase the platform flexibility of property in measure, analyze and use the KPI for supporting to decision. Moreover, the comprehension of huge diversity of KPI also contributes to increase a status of transparency, evolution and other aspects of ecosystem. Furthermore, an opportunity of important research is the identification, analysis and evaluation of KPI.

S5 presents a series of challenges and opportunities to future researches related to Open Innovation in SECO context. The first point highlighted by the authors, is the necessity of a better comprehension of the mechanisms for primary innovations. Moreover, according to the authors, it is necessary to understand the competitive architecture and organizational capacity that promote an open innovation, for example, investigate if some specific structures of ecosystem better support the software innovation than the rest. In addition, there is the necessity of obtain knowledge from studies about procedures and techniques that support innovation in SECO. The other important aspect of innovation in SECO is comprehend how SECO process and practices support the business innovation and software innovation combined and how the commercial agreements relative to relations and operations permit software innovation and collaboration across organizational boundaries.

Elements of MSECO are discussed in S17, however, communities and evangelists are indicated as elements least discussed in literature. Moreover, the authors affirm that more researches are necessary for a better comprehension of elements and its classifications. Also, highlight that social aspects are less explored in MSECO domain, what corroborates opportunity of research for SECO in a general manner (S1, S9, and S11). Studying users’ motivations of MSECO on shopping is an opportunity to research raised by S18.

Measure the ecosystem health would provide huge benefits for SECO industry and research, but there is difficult to define the measures in S2, S7 and S10. S7 and S10 still portray the challenge to define the concept of health in ECOs. For S10, other investigation areas about evaluate ecosystem health are found on an initial state, or even do not exist and, are necessary tools of data analysis, visualization and evaluation. The concern with quality is not specific to SECO, but the manner as quality is guaranteed has been pointed by S2 as a field that was not approached in an efficient manner in literature. Adoption of traditional methods of quality assurance cannot necessarily work in a SECO. On this way, there is a need of strategies to guarantee specific qualities of SECO, what open a field of several researches’ options. On the same line, S12 exposes that quality studies in SECO until now lacks investigations of specifics needs of different domains.

Related to gaps of research about requirements in SECO, S1 cites as a challenge overcome the techniques obstacles and socio-organizations for coordination and communication of requirements in projects geographically distributed. S2 and S9 present elicitation of requirements in SECO as an interesting challenge, due to parts interested are multiples and distant of ecosystems management. Moreover, it is necessary studies to search the equilibrium among the necessity of different parts interested. In S12 results, it is possible to observe that only one article addresses the requirement selection and there is not articles published about tools to analyze the requirements. Only one publication with notations and semantic for requirement modeling was found, but have not found articles about standards to specific the users’ requirement and tools to model the requirement. The validity of requirements and the activities of management are less addressed. The traceability is a significant task to provide changes on requirements. Therefore, for S12, more researches must be done in all activities of requirement engineer cited above. 

S1 presented research directions for SECO. Among them, S1 authors cited the need to determine what the best coping strategies in any ecosystem are. S2 portrayed that one field that has not been addressed in the literature is decision making in SECO. S15 defined a research agenda for governance in SECO and pointed out that there is a need for research to: (1) defining a common vocabulary in SECO governance, as several studies adopt related terms such as management and orchestration to refer-to-governance mechanisms; (2) analyzing the interaction between governance mechanisms and health measures, as health measures provide operational indicators on how SECO are governed; (3) the governance of developer ecosystems, because a better understanding of developer interests and behaviors is needed; (4) governance in OSSECO, considered as a promising line of research; and (5) understanding interactions between ecosystems, especially the interaction of large ECOS.

\subsection{The use of EvidenceSET tool}

\begin{figure}[h]
	\centerline{\includegraphics[scale=0.52]{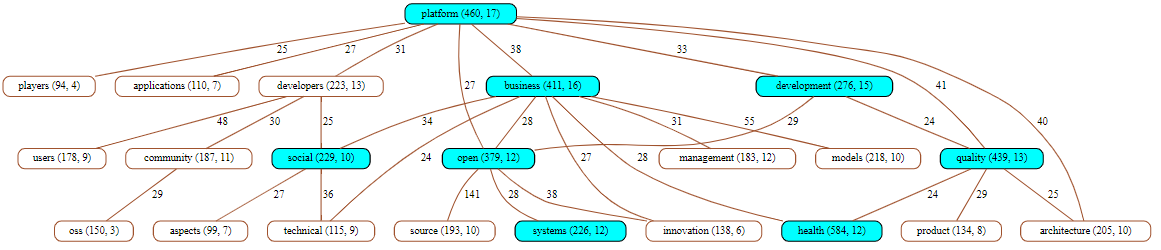}}
	\caption{EvidenceSET ``network graphviz" view.
	\label{fig:EvidencesetNetworkGraphviz}}
\end{figure}

As discussed above, we manually identified 1 high-order theme, 5 themes, 10 subthemes and dozens of categories. In addition, some relationships between all elements were identified. We modeled  the diagrams by using the Diagrams  tool after the processing of reading all studies, from oldest to newest one. We used EvidenceSET tool to support the identification of the elements. One of the EvidenceSET views that served as the unit of analysis for identifying themes is represented in the Figure \ref{fig:EvidencesetNetworkGraphviz}.

\begin{figure}[h]
	\centerline{\includegraphics[scale=0.6]{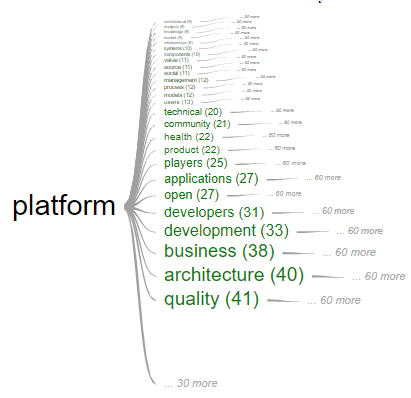}}
	\caption{EvidenceSET word tree view for the keyword ``platform".
	\label{fig:EvidencesetWordTreePlatform}}
\end{figure}

\begin{figure}[h]
	\centerline{\includegraphics[scale=0.6]{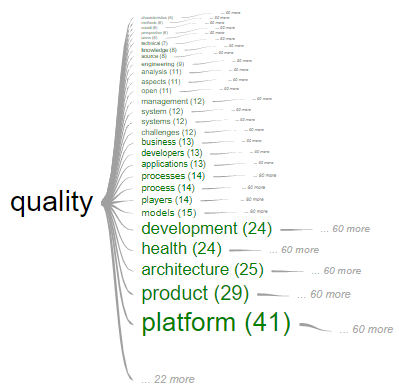}}
	\caption{EvidenceSET word tree view for the keyword ``quality".
	\label{fig:EvidencesetWordTreeQuality}}
\end{figure}

Figure \ref{fig:EvidencesetNetworkGraphviz} has 23 elements that according to the tool represent keywords identified in the secondary studies. In blue, they are the 'top 10' most frequent keywords. In white, the regular keywords. In parentheses, it is the ``keyword (x, y)" found, which x is the number of times that a specific ``keyword" was found, and y is the number of secondary studies that a ``keyword" was found. For instance, the word platform is the most frequent term considering the entire set of studies and it was found 460 times and in 17 of the 22 secondary studies. Furthermore, the connection between the terms has a weight that is given by the number of times that the linked terms occurred in the same sentence. For example, open and source. 141 means the number of times that both of keywords were found in the same statement. We observed that regardless of the concepts behind each term, it is possible to identify the SECO in the three-dimensional perspective with the existence of social, business and technical keywords. It is also possible to identify possible relationships between these concepts. Furthermore, it is possible, for instance, to identify possible strong relationships between terms such as health, platform and quality as visualized in Figure \ref{fig:EvidencesetNetworkGraphviz}, Figure \ref{fig:EvidencesetWordTreePlatform}, and Figure \ref{fig:EvidencesetWordTreeQuality}.

\begin{figure}[h]
	\centerline{\includegraphics[scale=0.6]{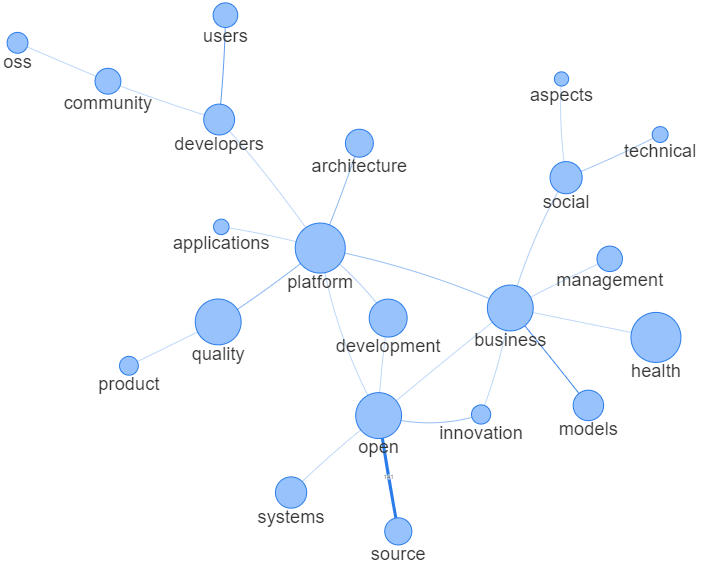}}
	\caption{EvidenceSET network visualization.
	\label{fig:EvidencesetNetworkGraph}}
\end{figure}

Figure \ref{fig:EvidencesetNetworkGraph} illustrates another EvidenceSET view: the network relationship between the keywords. The size of the circle of each keywords means the frequency of word citations and the different thicknesses of the links represent how strong the keywords are linked. It is important to note that while the themes are not necessarily mapped, their related concepts are. Then, through an interpretative analysis, the results were grouped in order to represent the themes shown in Figure \ref{fig:SECOThematicMap}, Figure 21, and the categories in Table \ref{tab:SubthemesAndCategories}.

\section{Discussion}
\label{sec:discussion}

This study identified that there is a growing interest at conducting both SLR and SMS studies in the context of SECO. We mapped that the secondary studies in the field of SECO has some peculiarities regarding the research method or design and the topics covered by the research or mapping questions used by the 22 systematics reviews. We detected that SECO appears in the three-dimensional perspective initially distinguished by Campbell and Ahmed (2010), evolved by Santos and Werner (2011a) and S1 has become a pattern to describe SECO. We observed the use of variations of these concepts among the following studies: S1, S2, S3, S4, S6, S8, S9, S12, S14, S15, S17, S18, S19, and S22. That is, 14 of the 22 secondary reviewers recognized the use of SECO in the three-dimensional perspective in their findings. S2, for instance, recognizes three main components of SECO, such as SECO Software Engineering, SECO Business and Management, and SECO Relations. Despite S2 did not mention directly technical, business or social terms, the definitions provided by this study are closely related to the dimensions proposed by Campbell and Ahmed (2010). The same occurs in S4, which specifies three dimensions of quality of OSS SECO: quality of platform (technical), quality of communities (social), and quality of networks (business). S14 used the dimensions evolved from Campbell and Ahmed (2010) when it grouped the measures for health ecosystems evaluation, an architecture named HEAL ME.

Concerning to SECO in the three-dimensional perspective, the technical dimension usually focuses at the software platform or the software product architecture from a perspective of domain engineering process, or SE, or architecture, and including aspects related to the development process. The business dimension usually efforts at the strategic planning, knowledge flow, organizational capabilities, decision-making, monitoring, intellectual property, value creation, governance, innovation, partnership models, and management aspects of the ecosystems. The social dimension emphases concepts as collaboration, social network opportunities, shared-knowledge, people engagement, people motivation, people realization of utility, interaction among actors, sense or ownership, promoting interactions among participants, promoting cultural change, and any others aspects resulted from the participation of the individuals, groups or the communities around the software platform or the software product architecture.

From the perspective of the thematic analysis, we have mapped these set of dimensions as a high-order theme in the research field of SECO, and each dimension of the SECO in the three-dimensional perspective being a theme of this high-order one. Some aspects linked to the themes were mapped as subthemes; the others linked to the subthemes, we identified as categories. Theses connections are illustrated in Figure 15 (used to represent the thematic analysis connections) and in Figure 20 (used to represent the three-dimensional perspective connections). The use of venn (in Figure 20) is not by chance . As identified from the analysis of the secondary studies, most of the time, an aspect or factor related to a dimension is not linked to just one dimension, it has implications for several others elements of an ecosystem. Thus, TS represents the technical and social perspectives together; BS, business and social; BT, business and technical, and BST, all of them. Accordingly, although we recognize the use of the SECO in the three-dimensional perspective, this relationship should not be seen in isolation, it must be interpreted in a holistic approach, given the number of implications to other dimensions. 

\begin{figure}[h]
	\centerline{\includegraphics[scale=0.3]{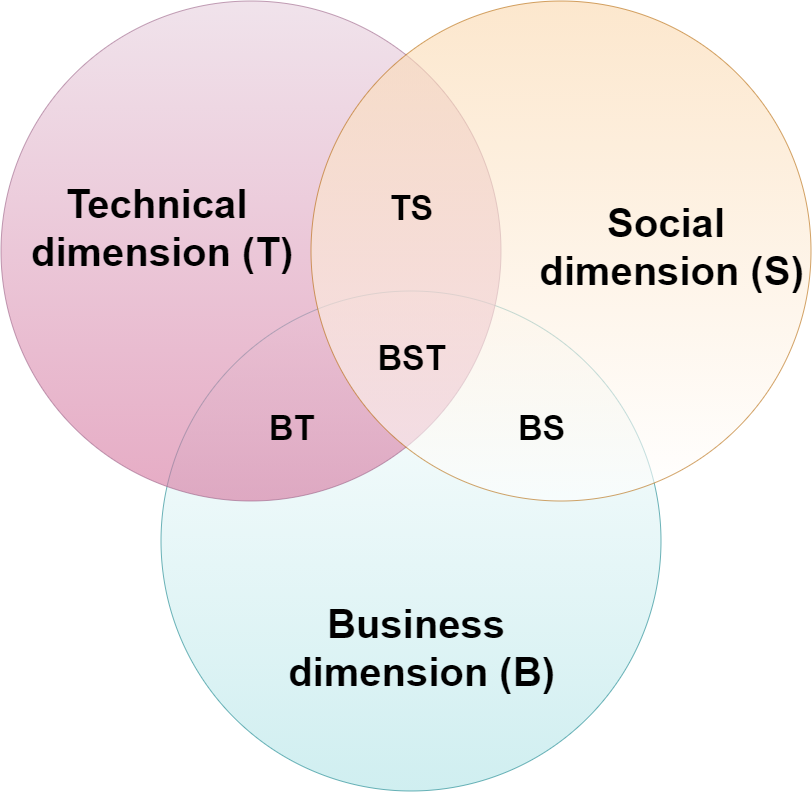}}
	\caption{SECO themes (SECO in the three-dimensional perspective) in a holistic approach.
	\label{fig:SECOThemes}}
\end{figure}

Complementing the traditional SECO in the three-dimensional perspective view, presented in Figure \ref{fig:SECOThemes}, two themes were identified in the thematic synthesis performed in this study and are inserted in this perspective, a theme related to the SECO assessment model (SEAM) and a management theme. To represent these 5 elements, a thematic model was defined and will be presented in the next subsection.

\subsection{A thematic model for SECO}

\begin{figure}[h]
	\centerline{\includegraphics[scale=0.6]{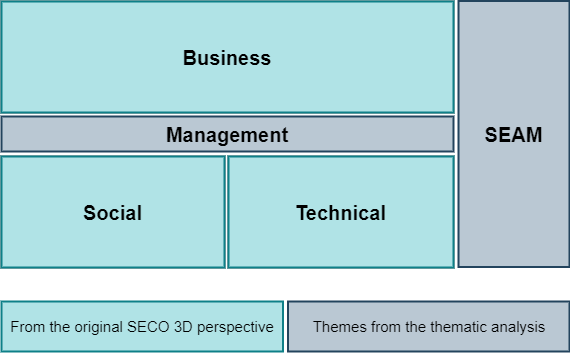}}
	\caption{A thematic model for SECO: social, technical, business, management and the SEAM.
	\label{fig:ThematicModelSECO}}
\end{figure}

To build the thematic model for SECO, illustrated in Figure \ref{fig:ThematicModelSECO}, the SECO in the three-dimensional perspective was taken into account, as well as the two themes found through the thematic synthesis performed by the analysis of the secondary studies included in this study. The first theme comes from the relation between the elements of the SECO in the three-dimensional perspective view, we named it as management. In this context, management is seen as the set of practices, principles, policies, procedures, skills, competencies, and processes that plan, build, deliver, and support software services and products in an SECO by aligning the business, technical, and social dimensions. In addition, management is tailored to the policy guidance and monitoring defined through ecosystem governance. Through the thematic analysis, it was also possible to observe a solid connection between the terms health, quality, evaluation, measures, and indicators in the SECO context. The analysis of these concepts allowed us to define a theme called SEAM. This theme is an integrated set of techniques, tools, practices, and methodologies used to specifically evaluate, through measures, any aspect associated with one or more of the SECO in the three-dimensional perspective.

\subsubsection{Management Theme}

Figure 21 illustrates a thematic model for SECO. In the middle of the image can be recognized a new aspect proposed for this model: the existence of the management. We proposed this theme as a link between the elements of SECO in the three-dimensional perspective. We get inspired by S2 and S8. These studies connect business, organizational and management aspects of SECO as a way to classify the primary studies in their SLRs. In another study, S22, it covers the connection between management and the technical dimension. The authors visualize management as a subset of the technical dimension by arguing that technical dimension encompasses both the processes of the management and development of the product. Therefore, these aspects discipline how solutions are collaboratively planned, evolved and released to customers. S11, in turn, mapped several management models and practices such as a framework for sustainable software ecosystem management; the management practices in technology and innovation management processes in software ecosystem; and a conceptual model of the collaboration management process in the OSS community. 

On the other hand, S10 identified models for ecosystems management that aim to achieve the SECO’s health. In this context, we emphasize the need of distinction between business and management, the distinction between technical and management, and the distinction between social and management. Noticeably, this distinction is not a segregation. This is inspired by ISO/IEC 38500:2008 that is intended to provide a set of principles for directors to use when evaluating, directing and monitoring the use of information technology (IT) in their organizations. ISO/IEC 38500 claims that while certain actions are deliberated by the organization's body of directors, others are taken by the management body. While the first is in the domain of governance, the other is in the management domain.

According to COBIT 2019, an IT governance framework that is based on ISO/IEC 38500:2008, in most enterprises, overall governance is the responsibility of the board of directors, under the leadership of the chairperson. On the other side, management is the responsibility of the executive management, under the leadership of the chief executive officer (CEO). Thus, this distinction is need because the management is in charge to plan, build, run, and monitor in alignment with the direction set by the governance body. The governance body provides direction that is set through prioritization and decision-making. Thus, using ISO/IEC 38500:2008 and COBIT 2019 as theoretical foundations and based on the results of this study, we identified that in SECO context, the keystones represent the governance body of SECO. Besides, direction actions are conducted by them while others are performed by other actors of ecosystems located at the management domain that are related to management system of policies, processes, and structures that support the governance of the SECO. Moreover, we recognize that in a SECO perspective, an orchestrator can act in both roles: governing the SECO or managing it. When it occurs, the orchestrator ensures that its business needs are applicable to the SECO, from the smallest, to the largest, regardless of purpose, design, software architecture, type of evolution adopted, organizational structure, or business partnership structure. 

In the context of SECO, management practices play an important role by connecting business directions, and their implications to both social and technical perspectives. The importance to distinguish management and governance concepts is identified by S15 that argues that several studies adopt related terms such as management and orchestration to refer to governance mechanisms. Furthermore, S15 emphasizes the need to establish a common glossary and conceptual framework that collects these definitions into ecosystem governance knowledge.  In this context, we see the management dimension as the group of practices, principles, policies, procedures, skills, competences and processes that plan, build, deliver and support services and software products in a SECO aligning the business, technical and social dimensions. In addition, management is indented to the policy guidance and monitoring set through ecosystem governance.

\subsubsection{SEAM Theme}

Complementing the traditional SECO in the three-dimensional perspective showed in Figure 20, we also identified a solid connection between the terms health, quality, evaluation, assessments, measures and indicators in the context of SECO field. In fact, SECO health and quality are two concepts widely used by secondary studies. They are used together because whenever one occurred, there was the other as if the two concepts had a close relationship. S1, S2, S4, S6, S9-S12, S14, S15, S18, and S20, for instance, have directly cited or used these concepts together. S1, in particular, recognizes quality as a concept that would need more research in the field of SECO. S1 argues that the quality concern is not SECO specific, but the way in which quality can be measured certainly is. In addition, they join the quality term in relation to OSS when they argue that there is a need to measure how quality can be measured per developer in an OSSECO. S2 recognized Quality Assurance (QA) as an area and it states that it has also not been efficiently addressed in the literature. It suggested that the adoption of traditional QA methods would not necessarily work in a SECO because of the separation of platform and actors. S4 proposed a quality model (QuESo) for OSSECO composed of two types of interrelated elements: quality characteristics and measures. The model covers three dimensions, as namely: 1) related to platform; 2) related to community; and 3) related to the ecosystem as a network of elements. S4 mapped several measure definitions with the same names that they were referred to in the SLR studies from where they were extracted. For instance, ecosystem connectedness is a measure highly connected to business dimension once is a property that keeps communities structure safe from risks, guaranteeing their well-being and health. Conversely, bug-tracking activity is strongly connected to the technical dimension, insofar as it represents the number of comments created in project bug tracker and total number of actions in the bug tracker.

S6 examines MSECO and brings several quality insights do mobile ecosystems. Some aspects discussed are as follows: the quality related to how to guarantee quality of both mobile applications and MSECO as a whole; specific QA practices for mobile applications because of the rapid evolution of a MSECO. As other studies, S6 also points out the need of more research regarding QA and MSECO. In this respect, S9 appears to extend one of future directions of S6 and it is fully devoted to QA once it investigated the challenges related to quality assurance in SECO. It defines QA as a set of activities carried out to ensure that the system has sufficient quality, and it argues that QA cannot be seen as an isolated activity, but rather as a set of activities that take place during different phases of the product’s life-cycle. The point is that one of the ways that the life-cycle can be defined is the standard ISO12207 (ISO/IEC/IEEE, 2008) for software life-cycle processes. It is the evidence that the field of SECO can make use of established theories to evolve specific characteristics related to ecosystems. S9 identifies what we also recognize in this study: product quality has a relationship with the much broader concept of ecosystem health used by many software ecosystem researchers. It recognizes that this is another term inherited from business ecosystem and has been the focus of specific studies as the framework proposed by Manikas and Hansen (2013b). 

S10 focused specifically on health of the software ecosystems and despite to other reviews they conclude that their findings show that the research field is quite immature, and there are few approaches and tools to support the evaluation work. In special, one of the research questions of S10 focused essentially on the SECO in the three-dimensional perspective. They have addressed which business, technical and social practices are used by existing approaches to evaluate the health of software ecosystems? S10 identified that governance is used to achieve health of ecosystems. S10 mapped that there are three big concepts that involves practices in accordance with health indicators defined by the work of Iansiti and Levien (2010): robustness, productivity, and niche creation. S10 also suggests that health evaluation is used for the process maturity evaluation for ecosystems; there are tens of measures defined by different authors, including S4. S11 focused on OSSM and states that the quality of OSSECO affects organizations, adopters, software developers and the OSSECO itself. This emphasizes the correlation between quality and the SECO in the three-dimensional perspective. As S2, S11 states that OSS quality models emerged due to the inability of traditional quality models to measure these unique OSS features. According to S11, OSSECO quality is quite different from the standard ones (e.g., ISO/IEC 25010). S11 also discuss about the need of monitoring a list of OSSECO quality sub-characteristics over time; and the link the gathered values with adopter needs by operationalizing quality requirements. S12 indicated that the most addressed topics on requirements engineering in SECO are identifying stakeholders and roles for requirements elicitation and requirements analysis. Regarding the quality, they found that there are two types of quality concerns: in-operation and in-development. 

A key point found on S14 is the use of the quality word to describe the context of the study: the quality of a SECO platform and its available products are important characteristics to ensure its success. S14 performed an evaluation process to application of health measures, and interact both of concepts together health and of quality. As other secondary studies, it states that the quality vision in this context goes beyond the traditional one of software development, for instance, addressing concepts such as SECO health. It proposed the HEAL ME architecture, which aims to perform the SECO health assessment automatically. This is an evidence of the correlation between the concepts of health, quality, measures and SECO assessments. In S15, the focus is on the health concept. It investigated the correlations between governance mechanisms and performance in the context of SECO. It identified 20 governance mechanisms and 52 measures and classified into the three health elements (productivity, robustness, niche creation). According to the S15, by selecting appropriate governance mechanisms, organizations can gain strategic advantage over others leading them to better performance and, consequently, to be healthier.

The word health in S19, and S20 is not related to the concept brought above, it relates to the health ecosystems in care and assistance. According to the S19, for instance, “health” is a keyword that mainly refers to the concept of the well-functioning of an ecosystem. It emphasizes the findings of this study by asserting that ecosystem health is employed as the ability of the ecosystem to endure and remain variable and productive over time. Thus, S19 research design concerned to isolate the type of ecosystem with the concept in order to limit the number of publications only to “ecosystems in healthcare”, and not “ecosystem health”. S22, finally, that focused on hackathons discusses how companies can leverage ecosystem health by conducting hackathons in which they identify 14 social, technical and business benefit that can be used as benefits from the perspective of ecosystem health. According to the authors, the overall health of a software ecosystem depends on the actions and decisions taken by each participant. They assess it by using three key measures inspired by the work of by the work of Iansiti and Levien (2010): robustness, productivity, and niche creation. 

According to the \cite{ISO15939}, measure can be defined as a variable to which a value is assigned as the result of measurement. The process of measuring can be related to standard of comparison, an estimate of what is to be expected for a person, organization, or situation of a SECO. At the time of defining SECO, S1 states that they are composed of several levels and each level has different research challenges starting from the effect of SECO architectural changes to developing general measures and measuring the SECO health. S1 also related the business dimension (or theme) to the process of defining, collecting and analyzing related measures of a SECO. Instead of the use of measure concept, S3 uses Key Performance Indicators (KPI) to assess whether and how well such objectives are met and what the platform owner can do to improve. S04 is fully devoted to measurement and provide a list of measure definitions of the quality of OSSECO. S7 mapped several studies that proposed or identified a correlation between health and measures of ecosystems. S10 suggests that the measures’ analysis can identify problems and improve SECO health. S14 proposed the HEAL ME architecture in SECO and used the effectiveness of the measures as evaluation process. In S15, the authors described a catalogue with a large variety measures to assess the health of software ecosystems.

Thus, these concepts suggest a solid connection between the terms health, quality, evaluation, assessments, measures and indicators in the context of SECO field. They are always related to their success characteristics regardless of the dimension considered, be it technical, business or social. This is an important finding, mainly in the planning phase of other secondary studies regarding the choices of keywords that will be used in the search strings. Combining the concepts above, we identified a theme named as Software Ecosystems Assessment Models (SEAM). We define this theme as an integrated set of techniques, tools, practices and methodologies used to evaluate, by the use of measures, any aspect associated to one or more of the SECO in the three-dimensional perspective.

\subsection{Recommendations and future research}	

From the analysis of the secondary studies included, we identified recommendations and future research for the SECO area. As a contribution, we outline some research opportunities:

\begin{itemize}
    \item We identified the need of further investigate the other derivations from the work of Moore’s (Moore 1993) and Iansiti and Levien (Iansiti and Levien 2004a; Iansiti and Levien 2004b). For instance, in a recent secondary study, Guggenberger et al. (2020) conducted a review in which they identified and other kinds of classification such as Platform Ecosystems, Service Ecosystems, and Innovation Ecosystems.
    \item Although this mapping identified different classifications, some ecosystem definitions are extremely similar to those found in SECO research field. For instance, the concept of Service Ecosystems is view as group of service providers, consumers, and composition developers that collaboratively create new services, thereby adding value to the service ecosystem. We believe that this definition could be applied to SECO or at least as a subtype of SECO, such as MSECO. Therefore, if there is already research recognizing different types of ecosystems and they all derive from the same source of business, there may be a benefit in studying other areas and incorporating the knowledge already produced.
    \item We have identified several categories and we grouped them along ten subthemes. For a deeper analysis of the relationships of these categories, we could have built a network map to visualize the connections between the subthemes and the themes, and between the subthemes and the categories. Nevertheless, it was out of scope of this study. Once it is done, it could produce a refinement of the subthemes and each category related to them, and make the links between them more explicit.
    \item We have identified and proposed through the thematic synthesis analysis, the existence of a theme that we called as management. We believe that more research could be done for a deeper understanding of distinction between governance and management in context of SECO. 
    \item We defined the SEAM theme as an integrated set of techniques, tools, practices, and methodologies used to specifically evaluate, through measures, any aspect associated with one or more of the SECO in the three-dimensional perspective. In this sense, we believe that the conduction of cases studies based on the characteristics of these tools and techniques could provide a more detailed characterization of its aspects and understand important limitations.   
    
\end{itemize}

\section{Limitations}
\label{sec:Limitations}

On this section will be presented potential threats to validity this study and possible limitations that occurred during its production, and our actions for its mitigation. The discussion is based on guidelines for addressing the threats to validity of secondary studies presented by \cite{ampatzoglou2019}. The threats on this context can be classified as study validity, data validity and research validity.

\subsection{Study Validity}

Initial steps of secondary studies can present some threats to the validity, mainly with regard to selection process of the studies. Thus, relevant studies cannot be included and irrelevant studies can be included due to a selection process with threats to validity. These threats can include: (1) limited source selection of researches; (2) ineffective search strings; and (3) skew on selection of secondary studies that were included. To mitigate a limitation (1), it was realized searches in eight digital libraries, including libraries that index publications of several publishers, journals and events, such as ACM Digital Library and Scopus. Related to limitation (2), the search string used was based on principals’ secondary studies published about SECO and in several synonyms related to types of secondary studies available in literature. It was used control studies on the process of string calibration, enable its sophistication. To mitigate the third type of potential threat (3), this research counted on the participation of four researchers, who are experienced on leading secondary studies and in SECO. Moreover, the criteria of inclusion/exclusion were applied independently by the first two authors, and validated by the two last. In divergences, consensus was sought between the parts.

\subsection{Data Validity}

Data extraction process and synthesis can cause a threat to data validity, because could induce to dubious results and conclusions with regarding, mainly, to: (1) no verified extraction of data; and (2) the researcher inclination during the data extraction process and synthesis. To mitigate the threat (1), all data transformations were documented, so that is possible to trace the synthesis until the secondary study correspondent through spreadsheet and concept mapping. Related to threats (2), we performed the data extraction independently by two researchers, after was done an analysis together and, finally, the data were condensed. After this process, the others two more experienced researches verified the result. Furthermore, as discussed above, we used an adaptation of the thematic synthesis process from the works of  \cite{connelly2016}, and \cite{cruzes2011b}. Then, to support, review and complement the identification of the elements, we used EvidenceSET tool. EvidenceSET is a web-based tool created by three of the authors. In this respect, there may be a bias to use the tool to confirm what the creators want to validate. To prevent or mitigate that bias, we used the tool only at the end of process of the theme identification. In this respect, all the themes were derived manually from the analysis of the secondary studies. When all the themes were identified, we applied the tool to complement the identification of themes and categories that could be missing. It was done by looking at the EvidenceSET views, once it provides a way to display relationships between concepts. Two of the authors performed this step, and then the other two reviewed the entire results. 

\subsection{Research Validity}

Ensure an auditable and replicable character for research is an important necessity for this type of study. However, threats to this necessity can show up when: (1) the tertiary study planning is not documented and/or followed; and (2) the changes done during execution of the study are not justified and mapped during the process. To mitigate the threats presented, we have defined a detailed protocol based on guidelines well established for systematic reviews of the literature \citep{petersen2015guidelines}.  Besides, we analyzed other tertiary studies published as reference to realization of this. Lastly, all necessary alterations that occurred during the production of this study were analyzed, justified and documented. 

\section{Conclusion}
\label{sec:conclusion}

SECO is the interaction, communication, cooperation, and synergy among a set of players such as organizations, systems, applications, platforms, vendors, developers, and end users. Depending on the type of interaction that actors have with others, each one can play a different role. Our tertiary study indicates a growth in the SECO community about conducting secondary studies. Publishing progress has increased considerably, especially in the years 2018-2019. We could identify that the studies concentrated around 10 areas or domains, as namely: general studies; quality and health concerns; MSECO; open innovation; OSS; software architecture; measures; partnership models; and requirements engineering. 

A preponderance of SMS is perceived over SLR. The first represents 63.64\% of publications (14), the second, 36.36\% (8). This may indicate that the secondary studies included focused more on broad reviews of specific SECO-related research topics. Thus, in general, the studies identified as much evidence as possible and classified the evidence found to structure the research area. Regarding the number of primary studies resulting from the screening process of secondary studies, we observed an average of 58 studies per study. However, this result was influenced by two secondary studies that have a significantly larger set primary studies than the others, the studies S5 and S8. If we disregard these two studies, the average drops to 39 primary studies per review. 

Regarding the types of publication, we could see that more than 50\% of the studies came from conferences, followed by a journal, workshop and one book chapter. The high number of studies published in conferences because the area of computing does not have a high number, when compared to other areas, of scientific journals. Besides this, there is still no journal specifically directed to the area. These facts lead many researchers to concentrate their efforts to publish in international conferences that gather the communities of the area.

In relation to quality analysis, we observed a score average of 5.02 for the group of the secondary studies analyzed, which ranged from a minimum score of 0.5 to a maximum of 8.5. We observed that the studies that came from workshops obtained the highest quality score index, regardless of whether it was for the type SLR or SMS. We observed that the quality score varies from year to year regardless of the number of studies published per year. Concerning to the countries of origin of the publications, we observed that the studies are concentrated in only 6 countries, as namely: Brazil, Denmark, Finland, Norway, Spain, and Sweden. 10 of the 22 studies came from authors in Brazil.

We identified that the SECO research filed comes from other types of ecosystems, in particular ECO, BECO, DBECO and DECO and influences on other kinds of derived ecosystems such as DATAECO. As main varieties of SECO, we could identify the following ones: Federated Embedded Systems Ecosystems (FESE), MSECO, OSSECO, and Technological health Ecosystems. We identified the need of SECO community to further investigate the results from others ecosystems, mainly from the DECO and DBECO communities. In fact, it was already done by S1, S4, S8, S9, S10, S11 and S15 at including Management of Emergent Digital EcoSystems (MEDES) as source of their research strategy or selecting primary studies or references from there. As pointed out by S7, we also believe that by including those references it would broaden the picture of others ecosystems concepts. In addition, we identify the need of further investigate the other derivations from the work of Moore’s \citep{moore1993predators} and \cite{iansiti2004, iansiti2004keystone}. For instance, in a recent secondary study, \cite{guggenberger2020} conducted a review in which they identified other kinds of classification such as Platform Ecosystems, Service Ecosystems, and Innovation Ecosystems. Although this mapping identified different classifications, some ecosystems definitions are extremely similar to those found in SECO research field. Therefore, we believe that if there is already research recognizing different types of ecosystems and they all derive from the same source of business, there may be a benefit in studying other areas and incorporating the knowledge already produced.

We identified that the majority of the authors have used three perspectives (regardless of the type of study – SLR or SMS - or topic of study) to summarize their findings, as namely: technical dimension, business dimension, and social dimension. Apparently, the SECO in the three-dimensional perspective is a consolidated approach to classify ecosystems in the SECO research field. With regard to thematic analysis, we could identify 1 high-order theme, 5 themes and 10 subthemes. We could observe that although some subthemes have a more business-oriented aspect, they always relate to the others. The same is true of the sub-themes that are more focused on social or technical perspective.  That is, a sub-theme in the SECO perspective should never be observed in isolation. Rather, it must be observed in a systemic manner with all its relationships and implications.

Regarding the research synthesis, we used a hierarchy of elements adapter from the works of \cite{connelly2016}, \cite{cruzes2011b}. We have mapped these set of dimensions as a high-order theme in the research field of SECO, and each dimension of the three-dimensional view being a theme of this high-order one. Some aspects linked to the themes can be seen as subthemes, and linked to the subthemes, we identify categories. The data extraction was performed starting from oldest to newest secondary study. By doing in that way, it was possible to identify all the knowledge used from secondary studies, despite the methodological content and the findings. The process of data extraction and themes identification and review took 6 stages. We could identify 1 high-order theme, 5 themes, 10 subthemes and dozens of categories. We highlight the opportunity to used methods and tool to support the extraction of data carried out in the thematic synthesis. This process had an important role in the process of analyzing relationships between items in the theme hierarchy covered in the Table \ref{tab:SubthemesAndCategories}. We believe that even though each theme can be seen in an isolation, the fact is that they are always interrelated with others concepts.

We expect that these results can help the SECO community in the process of analyzing themes and their connections mainly with other theories of other types of ecosystems derived from Moore’s \citep{moore1993predators} and \cite{iansiti2004, iansiti2004keystone}. As future work, we have identified the possibility of further exploring the relationships mapped in the thematic synthesis in terms of themes, subthemes and categories. In addition, we can analyze the themes identified in this research in case studies and conduct further research on the SEAM dimension.

\begin{acknowledgements}
This study was financed in part by the Coordenação de Aperfeiçoamento de Pessoal de Nível Superior – Brasil (CAPES) – Finance Code 001. The authors wish to acknowledge UNIRIO and FAPERJ (Proc. 211.583/2019) for the partial support to the work. The third author would like to thank the FAPEMA (Process BEPP-01608/21; UNIVERSAL00745/19). 
\end{acknowledgements}

\section*{Data Availability} 

The datasets generated during and/or analysed during the current study are available from the corresponding author on reasonable request.

\section*{Compliance with Ethical Standards}

The authors declared that they have no conflict of interest.

\appendix 
\clearpage

\section{Secondary Studies Inclued}
\label{SecStudies}

\begin{table}[h!]
\centering
\label{tab:SecondaryStudiesInclued}
\begin{tabular}{|l|l|l|l|}
\hline
\multicolumn{1}{|c|}{\textbf{ID}} & \multicolumn{1}{c|}{\textbf{Authors}} &
\multicolumn{1}{|c|}{\textbf{Title}} & \multicolumn{1}{c|}{\textbf{Year}} \\ \hline
S1  & \begin{tabular}[c]{@{}l@{}} Barbosa O, Santos R, Alves C,\\ Werner C, Jansen S \end{tabular}  & \begin{tabular}[c]{@{}l@{}} A systematic mapping study on software ecosystems\\ from a three-dimensional perspective \end{tabular} & 2013 \\ \hline
S2 & \begin{tabular}[c]{@{}l@{}} Manikas K, Hansen K \end{tabular} & Software ecosystems – A systematic literature review & 2013 \\ \hline
S3 & \begin{tabular}[c]{@{}l@{}} Fotrousi F, Fricker S,\\ Fiedler M, Le-Gall F \end{tabular} & \begin{tabular} [c]{@{}l@{}} KPIs for Software Ecosystems: A Systematic\\ Mapping Study \end{tabular} & 2014 \\ \hline
S4 & \begin{tabular}[c]{@{}l@{}} Franco-Bedoya O, Ameller D,\\ Costal D, Franch X \end{tabular} & \begin{tabular} [c]{@{}l@{}} Measuring the Quality of Open Source Software\\ Ecosystems Using QuESo \end{tabular} & 2015 \\ \hline
S5 & \begin{tabular}[c]{@{}l@{}} Papatheocharous E,\\ Andersson J, Axelsson J \end{tabular} & \begin{tabular} [c]{@{}l@{}} Ecosystems and Open Innovation for Embedded Systems:\\ A Systematic Mapping Study \end{tabular} & 2015 \\ \hline
S6 & \begin{tabular}[c]{@{}l@{}} Fontão A, Santos R,\\ Dias-Neto A \end{tabular} & \begin{tabular}[c]{@{}l@{}} Mobile Software Ecosystem (MSECO):\\ A Systematic Mapping Study \end{tabular} & 2015 \\ \hline
S7 & \begin{tabular}[c]{@{}l@{}} Hyrynsalmi S, Seppänen M,\\ Nokkala T, Suominen A,\\ Järvi A \end{tabular} & \begin{tabular} [c]{@{}l@{}} Wealthy, Healthy and/or Happy — What does\\ ‘Ecosystem Health’ Stand for? \end{tabular} &	2015 \\ \hline
S8 & Manikas K & \begin{tabular} [c]{@{}l@{}} Revisiting Software Ecosystems Research: A Longitudinal\\ Literature Study \end{tabular} & 2016 \\ \hline
S9 & Axelsson J, Skoglund M & \begin{tabular} [c]{@{}l@{}} Quality Assurance in Software Ecosystems: A Systematic\\ Literature Mapping and Research Agenda \end{tabular} & 2016 \\ \hline
S10	& \begin{tabular}[c]{@{}l@{}} Amorim S, Neto S,\\ McGregor J, Almeida E,\\ Chavez C \end{tabular} & \begin{tabular} [c]{@{}l@{}} How Has the Health of Software Ecosystems Been\\ Evaluated? A Systematic Review \end{tabular} & 2017 \\ \hline
S11 & \begin{tabular}[c]{@{}l@{}} Franco-Bedoya O, Ameller D,\\ Costal D, Franch X \end{tabular} & \begin{tabular} [c]{@{}l@{}} Open Source Software Ecosystems: A Systematic Mapping \end{tabular} & 2017 \\ \hline
S12 & \begin{tabular}[c]{@{}l@{}} Vegendla A, Duc A,\\ Gao S, Sindre G \end{tabular} & \begin{tabular} [c]{@{}l@{}} A Systematic Mapping Study on Requirements\\ Engineering in Software Ecosystems \end{tabular} & 2018 \\ \hline
S13 & \begin{tabular} [c]{@{}l@{}} García-Holgado A,\\ García-Peñalvo F \end{tabular} & \begin{tabular} [c]{@{}l@{}} Mapping the systematic literature studies about\\ software ecosystems \end{tabular} & 2018 \\ \hline
S14 & \begin{tabular}[c]{@{}l@{}} Carvalho I, Campos F, Braga R,\\ David J, Ströele V, Araújo M \end{tabular} & \begin{tabular} [c]{@{}l@{}} Health Evaluation in Software Ecosystems \end{tabular} & 2018 \\ \hline
S15 & \begin{tabular}[c]{@{}l@{}} Alves C, Oliveira J,\\ Jansen S \end{tabular} & \begin{tabular}[c]{@{}l@{}} Understanding Governance Mechanisms and Health in\\ Software Ecosystems: A Systematic Literature Review \end{tabular} & 2018 \\ \hline
S16 & Belo Í, Alves C & \begin{tabular}[c]{@{}l@{}} Partnership Models for Software Ecosystems: A\\ Systematic Mapping Study \end{tabular} & 2019 \\ \hline
S17 & \begin{tabular}[c]{@{}l@{}} Steglich, C, Marczak S,\\ Guerra L, Mosmann L,\\ Perin M, Figueira Filho F,\\ Souza, C \end{tabular} & \begin{tabular} [c]{@{}l@{}} Revisiting the Mobile Software Ecosystems Literature \end{tabular} &	2019 \\ \hline
S18 & \begin{tabular}[c]{@{}l@{}} Steglich C, Marczak S,\\ Souza C, Guerra L, Mosmann L,\\ Figueira Filho F, Perin M \end{tabular} & \begin{tabular} [c]{@{}l@{}} Social Aspects and How They Influence MSECO\\ Developers \end{tabular} & 2019 \\ \hline
S19	& \begin{tabular}[c]{@{}l@{}} Marcos-Pablos S,\\ García-Peñalvo F \end{tabular}  & \begin{tabular}[c]{@{}l@{}} Technological Ecosystems in Care and Assistance: A\\ Systematic Literature Review \end{tabular} & 2019 \\ \hline
S20	& \begin{tabular}[c]{@{}l@{}} García-Holgado A,\\ Marcos-Pablos S,\\ Therón R, García-Peñalvo J \end{tabular} & \begin{tabular} [c]{@{}l@{}} Technological Ecosystems in the Health Sector: a\\ Mapping Study of European Research Projects \end{tabular} & 2019 \\ \hline
S21 & \begin{tabular}[c]{@{}l@{}} Lima T, Werner C, Santos R \end{tabular} & \begin{tabular} [c]{@{}l@{}} Identifying Architecture Attributes in the Context of\\ Software Ecosystems Based on a Mapping Study \end{tabular} & 2019 \\ \hline
S22	& \begin{tabular}[c]{@{}l@{}} Valença G, Lacerda N,\\ Rebelo M, Alves C, Souza C \end{tabular} & \begin{tabular} [c]{@{}l@{}} On the Benefits of Corporate Hackathons for Software\\ Ecosystems – A Systematic Mapping Study \end{tabular} & 2019 \\ \hline
\end{tabular}
\end{table}

\clearpage


\bibliographystyle{spbasic}      % basic style, author-year citations
\bibliography{main}   % name your BibTeX data base

%\begin{thebibliography}{}
% etc
%\end{thebibliography}

\end{document}